\documentclass[prb,showpacs,groupedaddress]{revtex4}

\usepackage{amsmath}
\usepackage[dvips]{graphicx}

\begin{document}


\title{
Dynamical coherent-potential approximation approach to excitation
spectra in 3d transition metals\footnote{To be published in Phys. Rev. B.}
}



\author{Y. Kakehashi}
\email[]{yok@sci.u-ryukyu.ac.jp}
\affiliation{Department of Physics and Earth Sciences,
Faculty of Science, University of the Ryukyus,
1 Senbaru, Nishihara, Okinawa, 903-0213, Japan}

\author{M. Atiqur R. Patoary}
\affiliation{Department of Physics and Earth Sciences,
Faculty of Science, University of the Ryukyus,
1 Senbaru, Nishihara, Okinawa, 903-0213, Japan}

\author{T. Tamashiro}
\affiliation{Department of Physics and Earth Sciences,
Faculty of Science, University of the Ryukyus,
1 Senbaru, Nishihara, Okinawa, 903-0213, Japan}


\date{\today}
\vspace*{10mm}
\begin{abstract}
First-principles dynamical CPA (Coherent-Potential Approximation) 
for electron correlations has been developed further by taking into
account higher-order dynamical corrections with use of the asymptotic
approximation.  The theory is applied to the investigations
of a systematic change of excitation spectra in $3d$ transition metals 
from Sc to Cu at finite temperatures.  
It is shown that the dynamical effects damp main peaks in the densities 
of states (DOS) obtained by the local density approximation 
to the density functional theory, reduce the band broadening due to 
thermal spin fluctuations, create the Mott-Hubbard type bands in the 
case of fcc Mn and fcc Fe, and create a small hump corresponding to 
the `6 eV' satellite in the case of Co, Ni, and Cu.  Calculated DOS 
explain the X-ray photoelectron spectroscopy data as well as the 
bremsstrahlung isochromat spectroscopy data.  
Moreover, it is found that screening effects on the exchange energy
parameters are significant for 
understanding the spectra in magnetic transition metals. 
\end{abstract}

\pacs{71.20.Be,78.20.Bh,75.10.Lp,78.70.En}

\maketitle


\section{Introduction}


Electron correlations play an important role for understanding the
electronic structure, magnetism, metal-insulator transition , and the
high-temperature superconductivity in the solid-state physics, and 
thus a large number of
theories have been proposed to describe correlated electron
systems~\cite{fulde95}. 
Especially, in the case of magnetism, theories of electron 
correlations have been developed over fifty years to explain the 
ferromagnetism of transition metals, since the
Hartree-Fock approximation was recognized to overestimate their magnetic 
ordering energy. 

Gutzwiller~\cite{gutz63,gutz64,gutz65} 
proposed a variational theory which takes into account
on-site correlations by controlling the probability amplitudes of
doubly occupied states, and showed that electron correlations much
instabilize the ferromagnetism. Hubbard~\cite{hub63,hub64} 
developed a Green function method making use of the equation of 
motion method and a decoupling approximation
that leads to an alloy-analogy picture.  He succeeded in
describing the metal-insulator transition as well as the instability of
ferromagnetism due to electron correlations.
Kanamori~\cite{kana63} 
took into account the multiple scattering of electrons in the
low density limit, and showed that the effective Coulomb interaction for
the ferromagnetic instability is extremely renormalized by
electron-electron interactions. 

Above mentioned theories are limited to the ground state.  
Cyrot~\cite{cyrot72}
extended to finite temperatures an idea of alloy-analogy approximation 
for electron correlations proposed by Hubbard, on the basis of the
functional integral method~\cite{strat58,hub59,evan70,mora74}. 
He explained the T-P phase diagram for
metal-insulator transitions qualitatively.
Hubbard~\cite{hub79} and Hasegawa~\cite{hase79} 
independently developed the single-site spin 
fluctuation theory using the coherent potential approximation 
(CPA)~\cite{soven67,ehren76}.  
They showed that thermal
spin fluctuations much reduce the Curie temperatures obtained by the
Stoner theory for band calculations based on the local density
approximation (LDA) to the density functional theory~\cite{par89}.

The single-site spin fluctuation theory reduces to the Hartree-Fock one
at zero temperature because it is based on a high-temperature
approximation, {\it i.e.}, the static approximation to the
functional integral method.  
Therefore the theory does not take into account the
ground-state electron correlations as found by Gutzwiller, Hubbard, and
Kanamori. Kakehashi and Fulde~\cite{kake85} 
proposed a variational theory which adiabatically takes 
into account such correlations at finite
temperatures, and found further reduction of Curie temperature.  
Finally, Kakehashi~\cite{kake92} 
proposed the dynamical CPA which completely takes into account
the dynamical charge and spin fluctuations within the single-site
approximation, and clarified the dynamical effects on the momentum
distribution, magnetic
moment as well as excitation spectra using the Monte-Carlo technique.
In the next paper~\cite{kake02} 
which we refer to I in the following, we developed an
analytic method to the dynamical CPA, using the harmonic approximation.
In the recent paper~\cite{kake08} 
which we refer to II, we proposed the
first-principles dynamical CPA which
combines the dynamical CPA with the tight-binding
linear-muffintin-orbital (TB-LMTO)~\cite{ander94} 
base LDA+U Hamiltonian~\cite{anis97}.
Within the 2nd-order dynamical corrections to the static approximation,
we have shown that the dynamical CPA can describe the finite temperature
properties of excitations and magnetism in Fe and Ni quantitatively or
semiquantitatively. 

In this paper, we develop further the first-principles dynamical CPA by
taking into account higher-order dynamical corrections within an 
asymptotic approximation, and investigate a systematic change of
excitation spectra in 3$d$ transition metals from Sc to Cu.  
We will clarify the dynamical effects
on the excitation spectra in 3$d$ series at finite temperatures, and
will explain systematic change of X-ray photoelectron spectroscopy (XPS)
data~\cite{narm88} 
as well as the bremsstrahlung isochromat (BIS) data~\cite{spei84}.

Similar calculations have recently been 
performed at the ground state by Belashchenko {\it et. al.}~\cite{bela06} 
on the basis of the self-consistent local GW
approximation.  Their results of the first-principles
calculations, however, do not well describe the main peak
positions in the XPS and the BIS data, and seem to require further
development of the theory.

As we have proven in the separate papers~\cite{kake02-2,kake04}, 
the dynamical CPA is equivalent to the many-body CPA~\cite{hiro77}
developed in the disordered system, the dynamical mean field theory
(DMFT) in the metal-insulator transition in infinite 
dimensions~\cite{mull89,jarr92,geor93,geor96}, 
and the projection operator method CPA (PM-CPA)~\cite{kake04-1} 
for excitation problems.  
The dynamical CPA was
originally developed to describe the finite-temperature magnetism in
metallic systems starting from the static approximation
exact in the high-temperature limit.  The theory can 
treat the transverse spin fluctuations for arbitrary $d$ electron 
number at finite temperatures, though it is not easy in the
traditional quantum Monte-Carlo approach (QMC)~\cite{hir89}.
Moreover the theory
allows us to calculate excitation spectra up to the temperatures much
lower than those calculated by the QMC because the dynamical CPA is
an analytic theory which does not rely on the statistical
techniques. 

In the following section, we outline the first-principles dynamical CPA
presented in our paper II.
After having established the basic formulation, it is desired how to
calculate higher-order terms of individual harmonics in the dynamical part. 
In Sec. III, we calculate the higher-order terms
using asymptotic approximation, and obtain the expressions for
dynamical corrections.
In Sec. IV, we present the results of numerical calculations for 
the densities of
states (DOS) as the single-particle excitations in 3$d$
transition metals.  Calculations have been
performed at high temperatures, where the present approach works best. 
We will demonstrate that the dynamical CPA can explain a systematic 
change of the DOS in 3$d$ series from Sc to Cu.  
We also show that the screening effects on the exchange energy parameter
are significant for the description of the excitation spectra in Mn, 
Fe, and Co.  In the last section, we summarize the
present work and discuss future problems to be solved.

\section{First-principles TB-LMTO dynamical CPA}

We consider here a transition metal system with an atom per unit cell, 
and adopt the TB-LMTO Hamiltonian combined with a LDA+U Coulomb 
interactions as follows~\cite{kake08}.  
\begin{eqnarray}
H = H_{0} + H_{1} ,
\label{hhat}
\end{eqnarray}
\begin{eqnarray}
H_{0} = \sum_{iL\sigma} (\epsilon^{0}_{L} - \mu) \, \hat{n}_{iL \sigma} 
+ \sum_{iL jL^{\prime} \sigma} t_{iL jL^{\prime}} \, 
a_{iL \sigma}^{\dagger} a_{jL^{\prime} \sigma} \ ,
\label{h0}
\end{eqnarray}
\begin{eqnarray}
H_{1} = \sum_{i} 
\Big[ \sum_{m} U_{0} \, \hat{n}_{ilm \uparrow} \hat{n}_{ilm \downarrow} 
+ {\sum_{m > m^{\prime}}} 
(U_{1}-\frac{1}{2}J) \hat{n}_{ilm} \hat{n}_{ilm^{\prime}} -
{\sum_{m > m^{\prime}}} J   
\hat{\mbox{\boldmath$s$}}_{ilm} \cdot \hat{\mbox{\boldmath$s$}}_{ilm^{\prime}} 
\Big] \ . 
\label{h1}
\end{eqnarray}
Here $\epsilon^{0}_{L}$ in the noninteracting Hamiltonian $H_{0}$ 
is an atomic level on site $i$ and orbital $L$,
$\mu$ is the chemical potential, 
$t_{iL jL^{\prime}}$ is a transfer integral between orbitals $iL$ and 
$jL^{\prime}$. $L=(l,m)$ denotes $s$, $p$, and $d$ orbitals.
$a_{iL \sigma}^{\dagger}$ 
($a_{iL \sigma}$) is the creation (annihilation) operator for an
electron with orbital $L$ and spin $\sigma$ on site $i$, and 
$\hat{n}_{iL\sigma}=a_{iL \sigma}^{\dagger}a_{iL \sigma}$ is a charge
density operator for electrons with orbital $L$ and spin $\sigma$ on
site $i$. 
 
In the Coulomb interaction term $H_{1}$, we take into account on-site
interactions between $d$ electrons ($l=2$).
$U_{0}$ ($U_{1}$) and $J$ in $H_{1}$ are 
the intra-orbital (inter-orbital)
Coulomb and exchange interactions, respectively.  
$\hat{n}_{ilm}$ ($\hat{\mbox{\boldmath$s$}}_{ilm}$) with $l=2$ is 
the charge (spin)
density operator for $d$ electrons on site $i$ and orbital $m$.
It should be noted that the atomic level $\epsilon^{0}_{L}$ in 
$H_{0}$ is not identical with the LDA atomic level
$\epsilon_{L}$. The former is given by the latter as~\cite{anis97,anis97-2} 
$\epsilon^{0}_{L} = \epsilon_{L} - \partial
E^{U}_{\rm LDA}/\partial n_{iL\sigma}$. 
Here $n_{iL\sigma}$ is the charge density at the ground state, 
$E^{U}_{\rm LDA}$ is a LDA functional to the intraatomic Coulomb 
interactions.

The free energy of the system ${\mathcal F}$ is written in the
interaction representation as follows.
\begin{eqnarray}
e^{-\beta {\cal F}} = 
{\rm Tr} \left[ {\cal T} \exp \left( - \int^{\beta}_{0} (H_{0}(\tau) 
+ H_{1}(\tau)) \right) \right] .
\label{pf}
\end{eqnarray}
Here $\beta$ is the inverse temperature, ${\cal T}$ denotes the
time-ordered product for operators.  $H_{0}(\tau)$ 
($H_{1}(\tau)$) is the interaction representation of Hamiltonian 
$H_{0}$ ($H_{1}$).

We transform the interaction $H_{1}(\tau)$ in the free energy 
into a one-body dynamical potential $v$ making use of the 
Hubbard-Stratonovich transformation~\cite{mora74,kake08}.  
The transformation is a Gaussian formula for the
Bose-type operator $\{ b_{\mu} \}$.
\begin{eqnarray}
{\rm e}^{\displaystyle \ \sum_{mm^{\prime}} 
b_{m} \bar{A}_{mm^{\prime}} b_{m^{\prime}}} = 
\ \sqrt{\dfrac{{\rm det} \bar{A}}{\pi^{M}}} \int [ \prod_{m} dx_{m} ]
\ {\rm e}^{\displaystyle - \sum_{mm^{\prime}} (x_{m} \bar{A}_{mm^{\prime}} 
x_{m^{\prime}} - 2b_{m} \bar{A}_{mm^{\prime}}x_{m^{\prime}})} .
\label{gauss}
\end{eqnarray}
Here $\bar{A}_{mm^{\prime}}$ is a $M \times M$ matrix, and 
$\{ x_{m} \}$ are auxiliary field variables.
The above formula implies that the two-body interaction 
$\sum_{mm^{\prime}} b_{m} \bar{A}_{mm^{\prime}} b_{m^{\prime}}$ 
is transformed into a one-body interaction with a potential 
$- \sum_{m^{\prime}} 2 \bar{A}_{mm^{\prime}}x_{m^{\prime}}$ 
coupled with the random fields $\{ x_{m^{\prime}} \}$.

After making use of the transformation at each time $\tau$, the free energy 
${\cal F}$ is written in the Matsubara frequency representation as follows.
\begin{eqnarray}
e^{-\beta {\cal F}} = 
\int \Big[ \prod_{j=1}^{N} \prod_{m=1}^{2l+1} 
\delta \boldsymbol{\xi}_{jm} \delta \zeta_{jm} \Big]
\exp \big[ - \beta E[\boldsymbol{\xi},\zeta] \,\big] \ ,
\label{free1}
\end{eqnarray}
\begin{eqnarray}
E[\boldsymbol{\xi},\zeta] & = & - \beta^{-1} \ln {\rm Tr}
({\rm e}^{-\beta H_{0}}) 
- \beta^{-1} {\rm Sp} \ln (1-vg)   \nonumber \\ 
& & 
+ \frac{1}{4} \sum_{in} \sum_{mm^{\prime}} \left[ 
\zeta_{im}^{\ast}(i\omega_{n}) A_{mm^{\prime}} 
\zeta_{im^{\prime}}(i\omega_{n}) 
+ \sum_{\alpha=x,y,z} \xi^{\ast}_{im\alpha}(i\omega_{n})
B^{\alpha}_{mm^{\prime}} \xi_{im^{\prime}\alpha}(i\omega_{n})
\right] \ .
\label{exiz}
\end{eqnarray}
Here $N$ is the number of sites, 
$\zeta_{im}(i\omega_{n})$ 
($\xi_{im\alpha}(i\omega_{n})$) is the $n$-frequency component of 
an auxiliary field $\zeta_{im}(\tau)$ ($\xi_{im\alpha}(\tau)$) 
being conjugate with $i\hat{n}_{iL}(\tau)$
($\hat{m}_{iL\alpha}(\tau)=2\hat{s}_{iL\alpha}(\tau)$) for $l=2$. 
Sp in the second term at the r.h.s. (right-hand-side) of 
Eq. (\ref{exiz}) denotes a trace
over site, orbital, frequency, and spin.
$g$ is the temperature Green function for noninteracting system $H_{0}$.
The matrices $A_{mm^{\prime}}$ and $B^{\alpha}_{mm^{\prime}}$ 
are defined by 
$A_{mm^{\prime}} = U_{0}\delta_{mm^{\prime}} 
+ (2U_{1} - J)(1 - \delta_{mm^{\prime}})$, 
$B^{\alpha}_{mm^{\prime}} = J (1 - \delta_{mm^{\prime}})$ 
$(\alpha = x,y$), and $B^{z}_{mm^{\prime}} 
=  U_{0} \delta_{mm^{\prime}} + J (1 - \delta_{mm^{\prime}})$.

The functional integrals in Eq. (\ref{free1}) are defined by  
\begin{eqnarray} 
\int \Big[ \prod_{m=1}^{2l+1} \delta \zeta_{im} \Big]
= \int \prod_{m=1}^{N^{\prime}} 
\sqrt{\dfrac{\beta^{2l+1} {\rm det} A}{(4\pi)^{2l+1}}} 
\prod_{m=1}^{2l+1} d\zeta_{im}(0) \left[ \prod_{n=1}^{\infty}
\dfrac{\beta^{2l+1} {\rm det} A}{(4\pi)^{2l+1}}
d^{2}\zeta_{im}(i\omega_{n}) \right] \ .
\label{func2}
\end{eqnarray}
Here $d^{2}\zeta_{im}(i\omega_{n})=d {\rm Re}\zeta_{im}(i\omega_{n})
d {\rm Im}\zeta_{im}(i\omega_{n})$.  
The dynamical one-body potential $v$ at the r.h.s. of Eq. (\ref{exiz}) 
is defined by 
\begin{eqnarray}
(v)_{iLn\sigma jL^{\prime}n^{\prime}\sigma^{\prime}} 
= v_{jL\sigma\sigma^{\prime}}(i\omega_{n}-i\omega_{n^{\prime}})
\delta_{ij}\delta_{LL^{\prime}} \ ,
\label{dpot}
\end{eqnarray}
\begin{eqnarray}
v_{iL\sigma\sigma^{\prime}}(i\omega_{\nu}) = 
- \frac{1}{2}  \sum_{m^{\prime}} i A_{mm^{\prime}}
\zeta_{im^{\prime}}(i\omega_{\nu}) \delta_{l2}\delta_{\sigma\sigma^{\prime}} 
- \frac{1}{2} \sum_{\alpha} \sum_{m^{\prime}} 
B^{\alpha}_{mm^{\prime}} \xi_{im^{\prime}\alpha}(i\omega_{\nu})
\delta_{l2} 
(\sigma_{\alpha})_{\sigma\sigma^{\prime}} \ ,
\label{dpot2}
\end{eqnarray}
$\sigma_{\alpha}$ ($\alpha=x, y, z$) being the Pauli spin matrices.

In the dynamical CPA~\cite{kake02}, 
we introduce a site-diagonal coherent potential
\begin{eqnarray} 
(\Sigma)_{iLn\sigma jL^{\prime}n^{\prime}\sigma^{\prime}} =
\Sigma_{L\sigma}(i\omega_{n})
\delta_{ij}\delta_{LL^{\prime}}\delta_{nn^{\prime}}
\delta_{\sigma\sigma^{\prime}} \ ,
\label{sigma}
\end{eqnarray}
into the potential part of the energy functional 
$E[\boldsymbol{\xi},\zeta]$, and expand the correction $v-\Sigma$ 
with respect to sites.
\begin{eqnarray} 
E[\boldsymbol{\xi},\zeta] = N \tilde{\cal F}(\Sigma) 
+ \sum_{i} E_{i}[\boldsymbol{\xi}_{i},\zeta_{i}] + \Delta E \ .
\label{exiz2}
\end{eqnarray}
Here the zero-th order term $\tilde{\cal F}(\Sigma)$ is a coherent part 
of the free energy which is defined by
\begin{eqnarray} 
\tilde{\cal F}(\Sigma) 
= - (N\beta)^{-1} {\rm ln} {\rm Tr}({\rm e}^{-\beta H_{0}}) 
- (N\beta)^{-1} {\rm Sp} \ln (1 - \Sigma g) \ . 
\label{fcoh}
\end{eqnarray}
The next term in Eq. (\ref{exiz2}) is a sum of the single-site
energies $E_{i}[\boldsymbol{\xi}_{i},\zeta_{i}]$.
The dynamical CPA neglects the higher-order terms $\Delta E$
associated with inter-site correlations.

The free energy per atom is finally given by~\cite{kake02,kake08}
\begin{eqnarray}
{\mathcal F}_{\rm CPA} = \tilde{\mathcal F}[\Sigma]
- \beta^{-1} {\rm ln} \, \int \Big[ \prod_{\alpha} 
\sqrt{\dfrac{\beta \tilde{J}_{\alpha}}{4\pi}}
d \xi_{\alpha} \Big] \,
{\rm e}^{\displaystyle -\beta E_{\rm eff}(\mbox{\boldmath$\xi$})} .
\label{fcpa2}
\end{eqnarray}
Here $\tilde{J}_{x}=\tilde{J}_{y}=\tilde{J}_{\bot}=[1-1/(2l+1)]J$, 
$\tilde{J}_{z} = U_{0}/(2l+1) + \tilde{J}_{\bot}$, and
we expressed the single-site term (the second term at the
r.h.s. Eq. (\ref{fcpa2})) with use of an effective potential 
$E_{\rm eff}(\mbox{\boldmath$\xi$})$ projected onto a large static 
field variables $\xi_{\alpha} = \sum_{m} \xi_{m\alpha}(0)$.
Moreover we have omitted the site indices for simplicity.  

The effective potential 
$E_{\rm eff}(\mbox{\boldmath$\xi$})$ consists of 
the static contribution $E_{\rm st}(\mbox{\boldmath$\xi$})$ and 
the dynamical correction term 
$E_{\rm dyn}(\mbox{\boldmath$\xi$})$.
\begin{eqnarray}
E_{\rm eff}(\mbox{\boldmath$\xi$}) = E_{\rm st}(\mbox{\boldmath$\xi$}) 
+ E_{\rm dyn}(\mbox{\boldmath$\xi$}) .
\label{eeff}
\end{eqnarray}
The former is given as
\begin{eqnarray}
E_{\rm st}(\boldsymbol{\xi}) &=& 
- \dfrac{1}{\beta} \sum_{mn} 
{\rm ln} \Big[ 
(1 \! - \! \delta v_{L\uparrow}(0)F_{L\uparrow}(i\omega_{n}))
(1 \! - \! \delta v_{L\downarrow}(0)F_{L\downarrow}(i\omega_{n}))
- \dfrac{1}{4} \tilde{J}^{2}_{\bot} \xi^{2}_{\bot} 
F_{L\uparrow}(i\omega_{n})F_{L\downarrow}(i\omega_{n})
\Big]     \nonumber \\
&  & 
+ \dfrac{1}{4} \Big[
- (U_{0}-2U_{1}+J) \sum_{m} \tilde{n}_{L}(\boldsymbol{\xi})^{2}
- (2U_{1}-J) \tilde{n}_{l}(\boldsymbol{\xi})^{2}
+ \tilde{J}^{2}_{\bot} \xi^{2}_{\bot} + \tilde{J}^{2}_{z} \xi^{2}_{z}
\Big] .
\label{est2}
\end{eqnarray}
Here 
$\delta v_{L\sigma}(0) = v_{L\sigma}(0) - \Sigma_{L\sigma}(i\omega_{n})$, 
and $\xi^{2}_{\bot}= \xi^{2}_{x} + \xi^{2}_{y}$.  
$v_{L\sigma}(0)$ is a static potential given by 
$v_{L\sigma}(0) = [(U_{0}-2U_{1}+J)\tilde{n}_{lm}(\boldsymbol{\xi})+
(2U_{1}-J)\tilde{n}_{l}(\boldsymbol{\xi})]/2 -
\tilde{J}_{z}\xi_{z}\sigma/2$.
The electron number $\tilde{n}_{L}(\boldsymbol{\xi})$ for a given 
$\boldsymbol{\xi}$ is expressed by means of an impurity Green function 
as
\begin{eqnarray}
\tilde{n}_{L}(\boldsymbol{\xi}) = \frac{1}{\beta} \sum_{n\sigma}
 G_{L\sigma}(\boldsymbol{\xi}, i\omega_{n}) ,
\label{nlxi}
\end{eqnarray}
and $\tilde{n}_{l}(\boldsymbol{\xi}) = 
\sum_{m} \tilde{n}_{L}(\boldsymbol{\xi})$.
The impurity Green function 
$G_{L\sigma}(\boldsymbol{\xi}, i\omega_{n})$ 
has to be determined self-consistently.  The explicit expression will 
be given later (see Eq. (\ref{gimp})). 
 
The coherent Green function $F_{L\sigma}(i\omega_{n})$ in Eq. (\ref{est2})
is defined by
\begin{eqnarray}
F_{L\sigma}(i\omega_{n}) = [(i\omega_{n} - \mbox{\boldmath$H$}_{0} 
- \mbox{\boldmath$\Sigma$}(i\omega_{n}))^{-1}]_{iL\sigma iL\sigma} .
\label{fls}
\end{eqnarray}
Here $(\mbox{\boldmath$H$}_{0})_{iL\sigma jL^{\prime}\sigma}$ 
is the one-electron Hamiltonian matrix for the noninteracting
Hamiltonian $H_{0}$, and 
$(\mbox{\boldmath$\Sigma$}(i\omega_{n}))_{iL\sigma jL^{\prime}\sigma} = 
\Sigma_{L\sigma}(i\omega_{n})\delta_{ij}\delta_{LL^{\prime}}$.

The dynamical potential $E_{\rm dyn}(\mbox{\boldmath$\xi$})$
in Eq. (\ref{eeff}) has been obtained within the harmonic
approximation~\cite{kake02,kake08,amit71,dai91}. 
It is based on an expansion of $E_{\rm dyn}(\boldsymbol{\xi})$ with
respect to the frequency mode of the dynamical potential 
$v_{L\sigma\sigma^{\prime}}(i\omega_{\nu})$, where 
$\omega_{\nu}=2\nu\pi/\beta$.  The harmonic approximation is the neglect
of the mode-mode coupling terms in the expansion.  We have then 
(see Eq. (55) in our paper II) 
\begin{eqnarray}
E_{\rm dyn}(\boldsymbol{\xi}) = - \beta^{-1} {\rm ln} 
\left[ 1 + \sum^{\infty}_{\nu=1} \,(\overline{D}_{\nu} -1) \right] .
\label{edyn1}
\end{eqnarray}
Here the determinant $D_{\nu}$ is a contribution from a dynamical
potential $v_{L\sigma\sigma^{\prime}}(i\omega_{\nu})$ with frequency
$\omega_{\nu}$, and the upper bar denotes a Gaussian
average with respect to the dynamical charge and exchange field
variables, $\zeta_{m}(i\omega_{n})$ and 
$\xi_{m\alpha}(i\omega_{n})$ ($\alpha = x, y, z$).  

The determinant $D_{\nu}$ is expressed as~\cite{kake08} 
\begin{eqnarray}
D_{\nu} = \prod_{k=0}^{\nu-1} \left[ \prod_{m=1}^{2l+1} D_{\nu}(k,m)
\right] ,
\label{dnu2}
\end{eqnarray}
\begin{eqnarray}
D_{\nu}(k,m) = \left| 
\begin{array}{@{\,}ccccccc@{\,}}
\ddots &         &         &      &    &   &  \\
       & 1       & 1           &        & 0  & &  \\
       & a_{-\nu+k}(\nu,m) & 1            & 1    &    & &  \\
       &                & a_{k}(\nu,m) & 1        & 1  & &  \\
       &                &      & a_{\nu+k}(\nu,m) & 1  & 1 & \\
       & 0       &      &          & a_{2\nu+k}(\nu,m) & & \\
       &                &      &          &                & \ddots \ \ \  & \\
\end{array}
\right| \ .
\label{dnukm}
\end{eqnarray}
Note that 1 in the above determinant denotes the $2 \times 2$ unit matrix, 
$a_{n}(\nu,m)$ is a $2 \times 2$ matrix
defined by 
\begin{eqnarray}
a_{n}(\nu,m)_{\sigma\sigma^{\prime}} = 
\sum_{\sigma^{\prime\prime}\sigma^{\prime\prime\prime}
\sigma^{\prime\prime\prime\prime}} 
v_{L\sigma\sigma^{\prime\prime}}(i\omega_{\nu}) 
\tilde{g}_{L\sigma^{\prime\prime} \sigma^{\prime\prime\prime}}
(i\omega_{n}-i\omega_{\nu})
v_{L\sigma^{\prime\prime\prime}\sigma^{\prime\prime\prime\prime}}
(-i\omega_{\nu}) 
\tilde{g}_{L\sigma^{\prime\prime\prime\prime} \sigma^{\prime}}(i\omega_{n}) \ ,
\label{annum}
\end{eqnarray}
\begin{eqnarray}
\tilde{g}_{L\sigma\sigma^{\prime}}(i\omega_{n}) = [(F_{L}(i\omega_{n})^{-1} - 
\delta v_{0})^{-1}]_{\sigma\sigma^{\prime}} \ .
\label{gst}
\end{eqnarray}
Here $v_{L\sigma\sigma^{\prime}}(i\omega_{\nu})$ is defined by Eq. 
(\ref{dpot2}).
$\tilde{g}_{L\sigma \sigma^{\prime}}(i\omega_{n})$ is the impurity
Green function in the static approximation,
$(F_{L}(i\omega_{n}))_{\sigma\sigma^{\prime}} = 
F_{L\sigma}(i\omega_{n})\delta_{\sigma\sigma^{\prime}}$, and 
$\delta v_{0}$ is defined by 
$(\delta v_{0})_{\sigma\sigma^{\prime}} = 
v_{L\sigma\sigma^{\prime}}(0) - 
\Sigma_{L\sigma}(i\omega_{n})\delta_{\sigma\sigma^{\prime}}$.

The determinant $D_{\nu}(k,m)$ is expanded with respect to the 
dynamical potential as follows. 
\begin{eqnarray}
D_{\nu}(k,m) = 1 + D^{(1)}_{\nu}(k,m) + D^{(2)}_{\nu}(k,m) + \cdots ,
\label{dnukm2}
\end{eqnarray}
\begin{eqnarray}
D^{(n)}_{\nu}(k,m) = \sum_{\alpha_{1}\gamma_{1} \cdots \alpha_{n}\gamma_{n}}
v_{\alpha_{1}}(\nu,m)v_{\gamma_{1}}(-\nu,m) \cdots 
v_{\alpha_{n}}(\nu,m)v_{\gamma_{n}}(-\nu,m) 
\hat{D}^{(n)}_{\{ \alpha\gamma \}}(\nu,k,m) \ .
\label{dnnukm}
\end{eqnarray}
Here the subscripts 
$\alpha_{i}$ and $\gamma_{i}$ take 4 values $0$, $x$, $y$, and $z$,
and
\begin{eqnarray}
v_{0}(\nu,m) = - \dfrac{1}{2} i \sum_{m^{\prime}} A_{mm^{\prime}}
\zeta_{m^{\prime}}(i\omega_{\nu})\delta_{l2} \ ,
\label{v0num}
\end{eqnarray}
\begin{eqnarray}
v_{\alpha}(\nu,m) = - \dfrac{1}{2} \sum_{m^{\prime}} 
B^{\alpha}_{mm^{\prime}} \xi_{m^{\prime}\alpha}(i\omega_{\nu})\delta_{l2} \ , 
\hspace*{8mm} (\alpha=x,y,z) \ .
\label{vanum}
\end{eqnarray}
Note that the subscript 
$\{ \alpha\gamma \}$ of $\hat{D}^{(n)}_{\{ \alpha\gamma \}}(\nu,k,m)$ 
in Eq. (\ref{dnnukm}) denotes a set of 
$(\alpha_{1}\gamma_{1}, \cdots, \alpha_{n}\gamma_{n})$.
The frequency dependent factors 
$\hat{D}^{(n)}_{\{ \alpha\gamma \}}(\nu,k,m)$  
consist of a linear combination of $2n$ products of the static Green
functions.  Their first few terms are given in Appendix A of our paper 
II~\cite{kake08}.
Approximate expressions for higher-order terms will be given in the next
section. 

Substituting Eq. (\ref{dnukm2}) into Eq. (\ref{dnu2}) and taking the
Gaussian average, we reach
\begin{eqnarray}
E_{\rm dyn}(\boldsymbol{\xi}) = - \beta^{-1} {\rm ln} 
\left( 1 + \sum^{\infty}_{n=1} \sum^{\infty}_{\nu=1} 
\overline{D}_{\nu}^{(n)} \right) ,
\label{edyn2}
\end{eqnarray}
and
\begin{eqnarray}
\overline{D}^{(n)}_{\nu} = \dfrac{1}{(2\beta)^{n}} 
\sum_{\sum_{km} l(k,m)=n} \sum_{\{ \alpha_{j}(k,m)\} }
\sum_{\rm P}
\prod_{m=1}^{2l+1} \prod_{k=0}^{\nu-1}
\Bigg[ \Big( \prod_{j=1}^{l(k,m)} C^{\alpha_{j}(k,m)}_{mm_{\rm p}} \Big)
\hat{D}^{(l(k,m))}_{\{ \alpha\alpha_{{\rm p}^{-1}} \} }(\nu,k,m) \Bigg] .
\label{dnubarn}
\end{eqnarray}
Here each element of 
$\{ l(k,m)\}\, (k=0, \cdots, \nu-1; m=1, \cdots , 2l+1)$ has a value of 
zero or positive integer.
$\alpha_{j}(k,m)$ takes one of 4 cases $0$, $x$, $y$, and $z$.
$j$ denotes the $j$-th member
of the $(k,m)$ block with $l(k,m)$ elements. 
P denotes a permutation of a set $\{ (j,k,m) \}$; 
${\rm P} \{ (j,k,m) \} = \{ (j_{\rm p},k_{\rm p},m_{\rm p}) \}$.
$\alpha_{{\rm p}^{-1}}$ 
means a rearrangement of $\{ \alpha_{j}(k,m) \}$ according to 
the inverse permutation P${}^{-1}$.  
The coefficient $C^{\alpha}_{mm^{\prime}}$ in Eq. (\ref{dnubarn}) 
is a Coulomb interaction defined by 
\begin{eqnarray}
C^{\alpha}_{mm^{\prime}} = \begin{cases}
-A_{mm^{\prime}} & (\alpha=0) \\
B^{\alpha}_{mm^{\prime}}  & (\alpha=x,y,z) \ .
\end{cases}
\label{cdef}
\end{eqnarray}

The coherent potential can be determined by the stationary condition
$\delta \mathcal{F}_{\rm CPA}/\delta \Sigma = 0$.  
This yields the CPA equation as~\cite{kake08} 
\begin{eqnarray}
\langle G_{L\sigma}(\mbox{\boldmath$\xi$}, i\omega_{n}) \rangle 
= F_{L\sigma}(i\omega_{n}) \ .
\label{dcpa3}
\end{eqnarray}
Here $\langle \ \rangle$ at the l.h.s. (left-hand-side) 
is a classical average taken with respect to the
effective potential $E_{\rm eff}(\mbox{\boldmath$\xi$})$.
The impurity Green function is given by 
\begin{eqnarray}
G_{L\sigma}(\mbox{\boldmath$\xi$}, i\omega_{l}) = 
\tilde{g}_{L\sigma\sigma}(i\omega_{l}) + 
\dfrac{\displaystyle \sum_{n} \sum_{\nu} 
\frac{\delta \overline{D}^{(n)}_{\nu}}
{\displaystyle \kappa_{L\sigma}(i\omega_{l})
\delta \Sigma_{L\sigma}(i\omega_{l})}}
{\displaystyle 1+ \sum_{n} \sum_{\nu} \overline{D}^{(n)}_{\nu}} \ .
\label{gimp}
\end{eqnarray}
Note that the first term at the r.h.s. (right-hand-side) is the impurity 
Green function in the static approximation, which is given 
by Eq. (\ref{gst}). 
The second term is the dynamical corrections, and
$\kappa_{L\sigma}(i\omega_{l})= 1 - F_{L\sigma}(i\omega_{l})^{-2}
\delta F_{L\sigma}(i\omega_{l})/\delta \Sigma_{L\sigma}(i\omega_{l})$.

Solving the CPA equation (\ref{dcpa3}) self-consistently, we obtain 
the effective medium.  
The electron number on each orbital $L$ is then calculated from
\begin{eqnarray}
\langle \hat{n}_{L} \rangle = 
\dfrac{1}{\beta} \sum_{n\sigma} F_{L\sigma}(i\omega_{n}) \ .
\label{avnl}
\end{eqnarray}
The chemical potential $\mu$ is determined from the condition 
$n_{e} = \sum_{L} \langle \hat{n}_{L} \rangle$.
Here $n_{e}$ denotes the conduction electron number per atom.
The magnetic moment is given by
\begin{eqnarray}
\langle \hat{m}^{z}_{L} \rangle = 
\dfrac{1}{\beta} \sum_{n\sigma} \sigma F_{L\sigma}(i\omega_{n}) \ .
\label{avml}
\end{eqnarray}
In particular, the $l=2$ components of magnetic moment are expressed 
as
\begin{eqnarray}
\langle \hat{\boldsymbol{m}}_{l} \rangle = 
\langle \boldsymbol{\xi} \rangle \ . 
\label{avmd}
\end{eqnarray}
The above relation implies that the effective potential 
$E_{\rm eff}(\mbox{\boldmath$\xi$})$
is a potential energy for a local magnetic moment
$\mbox{\boldmath$\xi$}$.

\section{Higher-order dynamical corrections in asymptotic approximation}

In our previous paper II~\cite{kake08}, 
we took into account the dynamical corrections
up to the second order in Eqs. (\ref{edyn2}) and (\ref{gimp}).  
We will obtain higher-order
terms in this section using an asymptotic approximation.

We note that the coupling constants 
$B^{x}_{mm^{\prime}}=B^{y}_{mm^{\prime}}=J(1-\delta_{mm^{\prime}})$
are much smaller than $A_{mm^{\prime}}$ and $B^{z}_{mm^{\prime}}$
because $U_{0}$ and $U_{1} \gg J$.  
The latter condition is not necessarily satisfied for Sc and Ti.
But $J$ in these elements are small as compared with the $d$ band width.
Thus we neglect the transverse
potentials, $v_{x}(\nu, m)$ and $v_{y}(\nu, m)$ in the higher-order 
dynamical corrections. 
This approximation implies that $a_{n}(\nu, m)_{\sigma -\sigma}=0$.  
The determinant $D_{\nu}(k,m)$ in Eq. (\ref{dnu2}) is then written 
by the products of the single-spin components as
\begin{eqnarray}
D_{\nu}(k,m) = D_{\nu\uparrow}(k,m)D_{\nu\downarrow}(k,m) .
\label{dnu3}
\end{eqnarray}
Here $D_{\nu\sigma}(k,m)$ is defined by Eq. (\ref{dnukm}) in which 
the $2 \times 2$ unit matrices have been replaced by 1 ({\it i.e.}, 
$1 \times 1$ unit matrices), and the $2 \times 2$ matrices 
$a_{n}(\nu, m)$ have been replaced by the $1 \times 1$ matrices 
$a_{n}(\nu, m)_{\sigma\sigma}$.  The latter is now given by
\begin{eqnarray}
a_{n}(\nu, m)_{\sigma\sigma} = \sum^{0,z}_{\alpha,\gamma}
 v_{\alpha}(\nu,m) v_{\gamma}(-\nu,m) \hat{h}_{\alpha\gamma\sigma}
e_{n\sigma}(\nu,m) ,
\label{annums}
\end{eqnarray}
\begin{eqnarray}
e_{n\sigma}(\nu,m) = \tilde{g}_{L\sigma}(n-\nu)\tilde{g}_{L\sigma}(n) .
\label{ennum}
\end{eqnarray}
Here $\hat{h}_{\alpha\gamma\sigma} = \delta_{\alpha\gamma} 
+ \sigma(1-\delta_{\alpha\gamma})$, and we used a notation 
$\tilde{g}_{L\sigma}(n) = \tilde{g}_{L\sigma\sigma}(i\omega_{n})$ for
simplicity. 
 
With use of the Laplace expansion, the determinant $D_{\nu\sigma}(k,m)$
can be written as 
\begin{eqnarray}
D_{\nu\sigma}(k,m) = \tilde{D}_{1\sigma}(\nu,k,m)D_{1\sigma}(\nu,k,m) 
- a_{k}(\nu,m)_{\sigma\sigma} 
\tilde{D}_{2\sigma}(\nu,k,m)D_{2\sigma}(\nu,k,m) .
\label{dnuskm}
\end{eqnarray}
Here $D_{n\sigma}(\nu,k,m)$ and $ \tilde{D}_{n\sigma}(\nu,k,m)$ are
defined by
\begin{eqnarray}
D_{n\sigma}(\nu, k, m) = \left| 
\begin{array}{@{\,}cccccc@{\,}}
1       & 1           &        & 0  & &  \\
a_{n\nu+k}(\nu,m)_{\sigma\sigma} & 1            & 1    &    & &  \\
             & a_{(n+1)\nu+k}(\nu,m)_{\sigma\sigma} & 1        & 1  & &  \\
             &      & a_{(n+2)\nu+k}(\nu,m)_{\sigma\sigma} & 1  & 1 & \\
             &      &          &                & \ddots \ \ \ & \\
0            &      &          &                &        & \\
\end{array}
\right| \ , \ \ \ \ 
\label{dnsnum}
\end{eqnarray}
\begin{eqnarray}
\tilde{D}_{n\sigma}(\nu, k, m) = \left| 
\begin{array}{@{\,}cccccc@{\,}}
1       & 1           &        & 0  & &  \\
a_{-n\nu+k}(\nu,m)_{\sigma\sigma} & 1            & 1    &    & &  \\
             & a_{-(n+1)\nu+k}(\nu,m)_{\sigma\sigma} & 1        & 1  & &  \\
             &      & a_{-(n+2)\nu+k}(\nu,m)_{\sigma\sigma} & 1  & 1 & \\
             &      &          &                & \ddots \ \ \ & \\
0            &      &          &                &        & \\
\end{array}
\right| \ .
\label{tdnsnum}
\end{eqnarray}

As we have shown in Appendix A in our paper I, $D_{n\sigma}(\nu,k,m)$
and $ \tilde{D}_{n\sigma}(\nu,k,m)$ are expanded as follows.
\begin{eqnarray}
D_{n\sigma}(\nu,k,m) &=& \nonumber \\
& & \hspace{-20mm} 1 + \sum^{\infty}_{i=1} (-1)^{i} 
\sum^{\infty}_{l_{1}=n}\sum^{\infty}_{l_{2}=l_{1}+2} \cdots
\sum^{\infty}_{l_{i}=l_{i-1}+2} a_{l_{1}\nu+k}(\nu,m)_{\sigma\sigma}
a_{l_{2}\nu+k}(\nu,m)_{\sigma\sigma} \cdots  
a_{l_{i}\nu+k}(\nu,m)_{\sigma\sigma} , \ \ \ \ \ \ 
\label{dnsexp}
\end{eqnarray}
\begin{eqnarray}
\tilde{D}_{n\sigma}(\nu,k,m) &=& \nonumber \\
& & \hspace{-25mm} 1 + \sum^{\infty}_{i=1} (-1)^{i} 
\sum^{\infty}_{l_{1}=n}\sum^{\infty}_{l_{2}=l_{1}+2} \cdots
\sum^{\infty}_{l_{i}=l_{i-1}+2} a_{-l_{1}\nu+k}(\nu,m)_{\sigma\sigma}
a_{-l_{2}\nu+k}(\nu,m)_{\sigma\sigma} \cdots  
a_{-l_{i}\nu+k}(\nu,m)_{\sigma\sigma} . \ \ \ \ \ \ 
\label{tdnsexp}
\end{eqnarray}

Substituting Eq. (\ref{annums}) into Eq. (\ref{dnsexp}), we obtain
\begin{eqnarray}
D_{n\sigma}(\nu,k,m) &=&  \nonumber \\
& & \hspace{-10mm}
\sum^{\infty}_{i=1} 
\sum^{0,z}_{\alpha_{1}\gamma_{1}\cdots\alpha_{i}\gamma_{i}} \!\!\!
v_{\alpha_{1}}(\nu,m)v_{\gamma_{1}}(-\nu,m) \cdots 
v_{\alpha_{i}}(\nu,m)v_{\gamma_{i}}(-\nu,m)
\hat{D}^{(i)}_{n\sigma}(\{\alpha\gamma\}, \nu, k, m) \ ,  \ \ \ \ \ \ 
\label{dnsexp2}
\end{eqnarray}
\begin{eqnarray}
\hat{D}^{(i)}_{n\sigma}(\{\alpha\gamma\}, \nu, k, m) = 
(-1)^{i} \,\hat{h}_{\alpha_{1}\gamma_{1}\sigma} \cdots 
\hat{h}_{\alpha_{i}\gamma_{i}\sigma} A^{(i)}_{n\sigma}(\nu, k, m)  \ ,
\label{dins}
\end{eqnarray}
\begin{eqnarray}
A^{(i)}_{n\sigma}(\nu, k, m) = 
\sum^{\infty}_{l_{1}=n}\sum^{\infty}_{l_{2}=l_{1}+2} \cdots
\sum^{\infty}_{l_{i}=l_{i-1}+2} e_{l_{1}\nu+k\sigma}(\nu,m)
e_{l_{2}\nu+k\sigma}(\nu,m) \cdots  
e_{l_{i}\nu+k\sigma}(\nu,m) .
\label{ains}
\end{eqnarray}
In the same way, $\tilde{D}_{n\sigma}(\nu,k,m)$ is expressed by Eq. 
(\ref{dnsexp2}) in which $A^{(i)}_{n\sigma}(\nu, k, m)$ has been
replaced by
\begin{eqnarray}
\tilde{A}^{(i)}_{n\sigma}(\nu, k, m) = 
\sum^{\infty}_{l_{1}=n}\sum^{\infty}_{l_{2}=l_{1}+2} \cdots \!\!\!\!\!
\sum^{\infty}_{l_{i}=l_{i-1}+2} e_{-l_{1}\nu+k\sigma}(\nu,m)
e_{-l_{2}\nu+k\sigma}(\nu,m) \cdots  
e_{-l_{i}\nu+k\sigma}(\nu,m) .
\label{tains}
\end{eqnarray}

The quantities $A^{(i)}_{n\sigma}(\nu, k, m)$ and
$\tilde{A}^{(i)}_{n\sigma}(\nu, k, m)$  contain the $i$-fold
summations.  In order to reduce these summations, we make use of an
asymptotic approximation.  The approximation is based on a
high-frequency behavior of $\tilde{g}_{L\sigma}(n)$ as
\begin{eqnarray}
\tilde{g}_{L\sigma}(n) = 
\dfrac{1}{i\omega_{n} - \epsilon^{0}_{L} + \mu - v_{L\sigma}(0)}
+ O\left(\dfrac{1}{(i\omega_{n})^{3}} \right) .
\label{gasym}
\end{eqnarray}
Then the product of $\tilde{g}_{L\sigma}(n-\nu)$ and 
$\tilde{g}_{L\sigma}(n)$ in $e_{n\sigma}(\nu, m)$ is written by their
difference as
\begin{eqnarray}
e_{n\sigma}(\nu, m) \sim 
\overline{q}_{\nu} (\tilde{g}_{L\sigma}(n-\nu) - \tilde{g}_{L\sigma}(n))
\ ,
\label{easym}
\end{eqnarray}
where $\overline{q}_{\nu}=\beta/2\pi\nu i$.

Substituting Eq. (\ref{easym}) into Eq. (\ref{ains}) successively, we
find
\begin{eqnarray}
A^{(i)}_{n\sigma}(\nu, k, m) \sim \,\frac{1}{i \,!} \,\overline{q}_{\nu}^{\,i}
\,\tilde{g}_{L\sigma}((n-1)\nu+k)\tilde{g}_{L\sigma}(n\nu+k) \cdots 
\tilde{g}_{L\sigma}((n+i-2)\nu+k) \ .
\label{aasym}
\end{eqnarray}
In the same way, we have
\begin{eqnarray}
\tilde{A}^{(i)}_{n\sigma}(\nu, k, m) \sim \,\frac{1}{i \,!} 
\,\overline{q}_{\nu}^{\,i}
\,\tilde{g}_{L\sigma}(-n\nu+k)\tilde{g}_{L\sigma}(-(n+1)\nu+k) \cdots 
\tilde{g}_{L\sigma}(-(n+i-1)\nu+k) \ .
\label{taasym}
\end{eqnarray}

Substituting $D_{n\sigma}(\nu,k,m)$ with Eq. (\ref{aasym})
and $\tilde{D}_{n\sigma}(\nu,k,m)$ with Eq. (\ref{taasym}) into 
Eq. (\ref{dnuskm}), we obtain
\begin{eqnarray}
D_{\nu\sigma}(k,m) = \sum^{\infty}_{l=1} 
\sum^{0,z}_{\alpha_{1}\gamma_{1}\cdots\alpha_{l}\gamma_{l}}
v_{\alpha_{1}}(\nu,m)v_{\gamma_{1}}(-\nu,m) \cdots 
v_{\alpha_{l}}(\nu,m)v_{\gamma_{l}}(-\nu,m)
\hat{D}^{(l)}_{\{\alpha\gamma\}\sigma}(\nu, k, m) \ ,
\label{dnusexp3}
\end{eqnarray}
\begin{eqnarray}
\hat{D}^{(l)}_{\{\alpha\gamma\}\sigma}(\nu, k, m) =
\Lambda^{(l)}_{\sigma}(\{\alpha\gamma\}) 
\frac{\overline{q}_{\nu}^{\,i}}{l\,!} B^{(l)}_{\sigma}(\nu, k, m)
\ .
\label{dlags}
\end{eqnarray}
Here
\begin{eqnarray}
\Lambda^{(l)}_{\sigma}(\{\alpha\gamma\}) = 
\begin{cases}
\ 1 & (\sigma=\uparrow) \\
(-1)^{l-n_{l}(\{\alpha\gamma\})} & (\sigma=\downarrow)
\end{cases}
\ ,
\label{lambda}
\end{eqnarray}
\begin{eqnarray}
B^{(l)}_{\sigma}(\nu, k, m) &=& 
\Big[ \prod^{l-1}_{j=0} \tilde{g}_{L\sigma}(j\nu+k) \Big] \nonumber \\
& & + \sum^{l-1}_{i=0} \dfrac{(-)^{l-i} l!}{i! (l-i)!}
\Big[ \prod^{i-1}_{j=-(l-i)} \tilde{g}_{L\sigma}(j\nu+k) \Big]
\Big[ 1 + \dfrac{l-i}{\overline{q}_{\nu}^{\,i}} 
\tilde{g}_{L\sigma}(i\nu+k) \Big] \ ,
\label{bls}
\end{eqnarray}
and $\hat{D}^{(0)}_{\{\alpha\gamma\}\sigma}(\nu, k, m) = 1$.
$n_{l}(\{\alpha\gamma\})$ in Eq. (\ref{lambda}) 
is the number of $\{\alpha_{i}\gamma_{i}\}$ 
pairs such that $\alpha_{i}=\gamma_{i}$ among the $l$ pairs.

Substituting Eq. (\ref{dnusexp3}) into Eq. (\ref{dnu3}), we obtain 
$\hat{D}^{(n)}_{\{\alpha\gamma\}}(\nu,k,m)$ in the asymptotic approximation.
\begin{eqnarray}
\hat{D}^{(n)}_{\{\alpha\gamma\}}(\nu, k, m) =
\sum^{n}_{l=0} \hat{D}^{(l)}_{\{\alpha_{1}\gamma_{1}
\cdots \alpha_{l}\gamma_{l}\}\uparrow}(\nu, k, m)
\hat{D}^{(n-l)}_{\{\alpha_{l+1}\gamma_{l+1}
\cdots \alpha_{n}\gamma_{n}\}\downarrow}(\nu, k, m) \ .
\label{dnag2}
\end{eqnarray}
Here we wrote the subscript at the r.h.s. explicitly to avoid
confusion.  Note that the values of $\alpha_{i}$ and $\gamma_{i}$ are
limited to $0$ or $z$ in the present approximation.
When there is no orbital degeneracy, Eq. (\ref{bls}) reduces to the
result of the zeroth asymptotic approximation in our paper 
I~\cite{kake02}.

In the actual applications we make use of the exact form up to a certain
order of expansion in $\overline{D}^{(m)}_{\nu}$, 
and for higher order terms we adopt an approximate
form (\ref{dnag2}).  
In this way, we can take into account dynamical corrections
systematically starting from both sides, the weak interaction limit 
and the high-temperature one.

\section{Numerical results of excitation spectra}

In the numerical calculations, we took into account the dynamical
corrections up to the second order ($n \le 2$) exactly, 
and the higher-order terms
up to the fourth order within the asymptotic approximation.
Summation with respect to $\nu$ in Eqs. (\ref{edyn2})
and (\ref{gimp}) was taken up to $\nu = 100$ for $n=1$ and
$2$, and up to $\nu = 2$ for $n = 3, \ 4$.

When we solved the CPA equation (\ref{dcpa3}), we adopted a decoupling 
approximation to the thermal average of impurity Green 
function~\cite{kake81}, 
{\it i.e.}, 
\begin{eqnarray}
\langle G_{L\sigma}(\xi_{z}, \xi^{2}_{\perp}, i\omega_{n}) \rangle = 
\sum_{q=\pm} \frac{1}{2}
\left( 1 + q \dfrac{\langle \xi_{z} \rangle}
{\sqrt{\langle \xi^{2}_{z} \rangle}} \right)
G_{L\sigma}(q\sqrt{\langle \xi^{2}_{z} \rangle}, 
\langle \xi^{2}_{\perp} \rangle, i\omega_{n}) \ .
\label{gapprox}
\end{eqnarray}
Here we wrote the static exchange field $\boldsymbol{\xi}$ as 
$(\xi_{z}, \xi^{2}_{\perp})$ so that the decoupling approximation we
made becomes clearer.
The approximation is correct up to the second moment ({\it i.e.},
$\langle \xi^{2}_{\alpha} \rangle$) and allows us to 
describe the thermal spin fluctuations in a simple way.

On the other hand, we adopted a diagonal approximation~\cite{kirk70,korr58} 
to the coherent Green function at the r.h.s. of Eq. (\ref{dcpa3}). 
\begin{eqnarray}
F_{L\sigma}(n) = \int \dfrac{\rho^{\rm LDA}_{L}(\epsilon) d \epsilon}
{i\omega_{n} - \epsilon - \Sigma_{L\sigma}(i\omega_{n}) - 
\Delta\epsilon_{L}} \ .
\label{cohg2}
\end{eqnarray}
Here $\rho^{\rm LDA}_{L}(\epsilon)$ is the local density of states for the LDA
band calculation, and 
$\Delta \epsilon_{L} = (\epsilon_{L}-\epsilon^{0}_{L})\delta_{l2}$.
The approximation partly takes into account the effects of hybridization
between different $l$ blocks in the nonmagnetic state, but neglects the
effects via spin polarization.

The CPA equation with use of the decoupling approximation (\ref{gapprox})
yields an approximate solution to the full CPA equation (\ref{dcpa3}).  
For the calculations of excitation spectra, one needs more accurate
solution for the CPA self-consistent equation.  
We thus adopted the following average $t$-matrix
approximation~\cite{korr58,ehren76} (ATA)
after we solved Eq. (\ref{dcpa3}) with the decoupling approximation.  
\begin{eqnarray}
 \Sigma^{\rm ATA}_{L\sigma}(i\omega_{n}) =  \Sigma_{L\sigma}(i\omega_{n}) + 
\dfrac{\langle G_{L\sigma}(\xi_{z}, \xi^{2}_{\perp}, i\omega_{n})
\rangle -F_{L\sigma}(i\omega_{n})}
{\langle G_{L\sigma}(\xi_{z}, \xi^{2}_{\perp}, i\omega_{n}) \rangle
F_{L\sigma}(i\omega_{n})} \ .
\label{ata}
\end{eqnarray}
Here the coherent potential in the decoupling
approximation is used at the r.h.s., but the full average 
$\langle \ \rangle$ of the 
impurity Green function is taken.
The ATA is a one-shot correction to the full CPA (\ref{dcpa3}).
%
%
\begin{figure}
\includegraphics{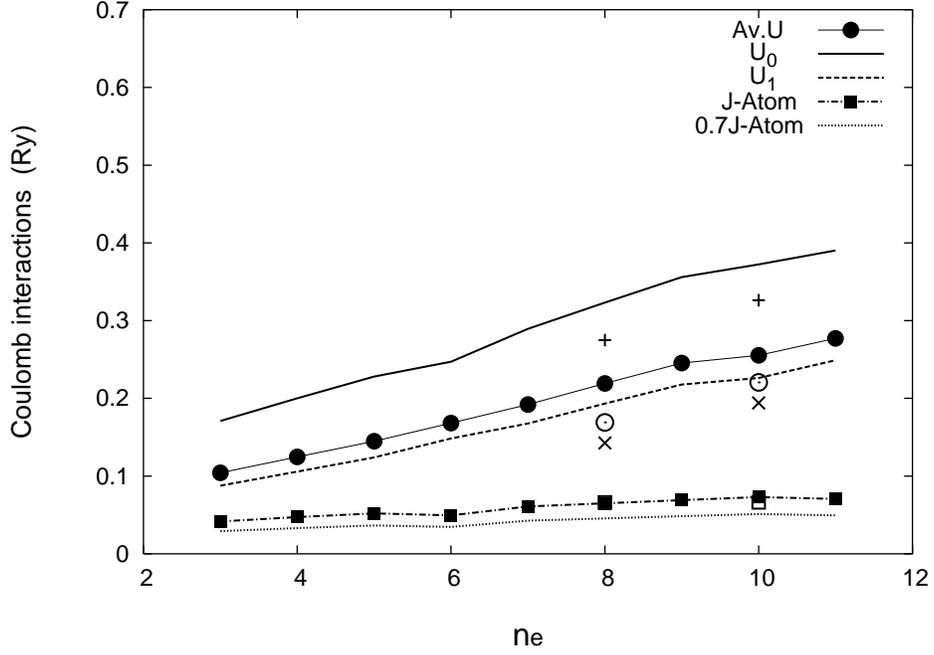}%
\caption{\label{figuj}
Intraatomic Coulomb and exchange energy parameters as a function of 
the conduction electron number of 3$d$ transition metals.
These parameters are obtained from the
band~\cite{bdyo89} and atomic~\cite{mann67} calculations.  
Averaged Coulomb interactions 
$\overline{U}$: closed circles and thin curve,
intraorbital Coulomb interactions $U_{0}$: solid curve,
interorbital Coulomb interactions $U_{1}$: dashed curve,
Hartree-Fock exchange interactions $\overline{J}$: 
closed squares and dot-dashed curve, 
and screened exchange interactions $0.7J$: dotted curve.
$\overline{U}$, $U_{0}$, $U_{1}$, $\overline{J}$ recommended by 
Anisimov {\it et. al.}~\cite{anis97-2} are
plotted by $\bigcirc$, $+$, $\times$, and $\sqcap \!\!\!\! \sqcup$, 
respectively,
for Fe ($n_{e}=8$) and Ni ($n_{e}=10$). 
}
\end{figure}
%

The coherent potential $\Sigma_{L\sigma}(z)$ on the real axis $z=\omega
+ i\delta$ is then calculated by using the Pad\'{e}
numerical analytic continuation method~\cite{vidberg77}. 
Here $\delta$ is an infinitesimal positive number.
The densities of states (DOS) as the single-particle excitations, 
$\rho_{L}(\omega)$ are 
calculated from the relation,
\begin{eqnarray} 
\rho_{L}(\omega) = - \frac{1}{\pi} \, {\rm Im} \, F_{L\sigma}(z) \ .
\label{dos}
\end{eqnarray}

We adopted the same lattice constants and structures as used by Andersen
{\it et. al.}~\cite{ander94} in order to investigate a systematic change
of excitations.  
For fcc Fe, we used the lattice constant 6.928
a.u. being observed at 1440 K.
The LDA calculations have been performed with use of the Barth-Hedin
exchange-correlation potential to make the TB-LMTO Hamiltonian (\ref{h0}).
In the present work all the dynamical CPA calculations have been 
performed at 2000 K in the paramagnetic state.

We adopted average Coulomb interaction parameters $\overline{U}$
obtained by Bandyopadhyav {\it et. al.}~\cite{bdyo89}, 
and the average exchange
interactions $\overline{J}$ obtained from the Hartree-Fock atomic
calculations~\cite{mann67}. 
The intra-orbital Coulomb interaction $U_{0}$, inter-orbital Coulomb 
interaction $U_{1}$, 
and the exchange interaction energy parameter $J$ were calculated from 
$\overline{U}$ and $\overline{J}$ as 
$U_{0} = \overline{U} + 8\overline{J}/5$, 
$U_{1} = \overline{U}-2\overline{J}/5$, and $J=\overline{J}$, 
using the relation $U_{0} = U_{1}+2J$.  

Calculated Coulomb interactions from Sc to Cu are plotted in Fig. 1 as 
a function of the valence electron number $n_{e}$.
For Fe and Ni, we adopted the values used 
by Anisimov {\it et. al.}~\cite{anis97-2},
which are also shown in the figure.  
Recent calculations suggest that the exchange interactions in
the 3$d$ metals are reduced by about 30\% as compared with their atomic
values~\cite{miya08}.  
These values are also shown in Fig. 1 by dotted line.
We will discuss the screening effects of $J$ on the spectra using 
the values.
%
\begin{figure}
\includegraphics{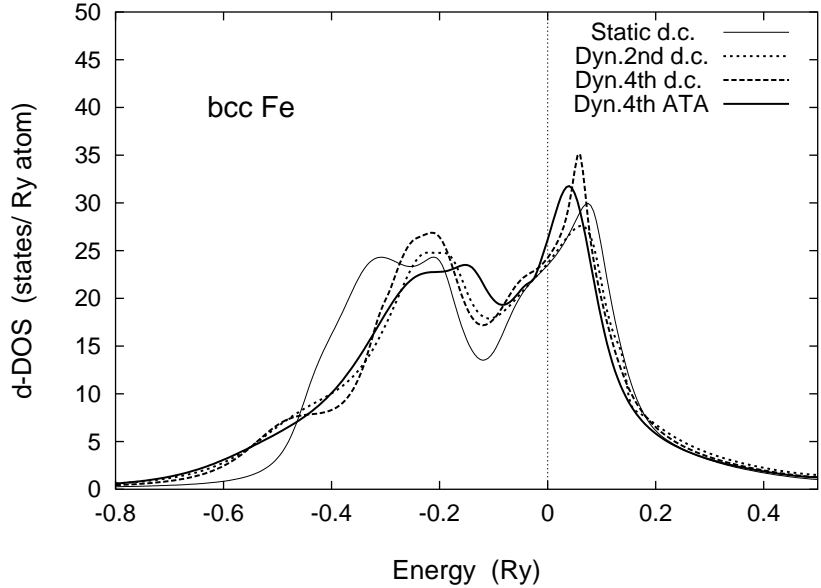}%
\caption{\label{figddos}
The $d$ partial densities of states (DOS) of bcc Fe at 2000K 
in various approximations.
The DOS in the static approximation with the decoupling scheme 
(\ref{gapprox}): thin solid curve, the DOS with the 2nd-order dynamical 
corrections and the decoupling (\ref{gapprox}): dotted curve, 
the DOS with the 4th-order dynamical corrections and the decoupling 
(\ref{gapprox}), and the DOS with the 4th-order dynamical 
corrections in the ATA.
}
\end{figure}
%

Before we present the results of excitation spectra in 3$d$ series, 
we briefly discuss the 4th-order dynamical effects.  
Figure 2 shows the $d$ partial DOS for
the paramagnetic bcc Fe on various levels of approximations.
The static approximation with the decoupling scheme (\ref{gapprox}) 
causes too strong thermal spin
fluctuations with large exchange splitting, and yields the
two-peak structure as shown by a thin curves in Fig. 2.  The second
order dynamical corrections suppress the thermal spin fluctuations and 
reduce the $d$ band width as well as the dip
at $\omega=-0.12$ Ry.  The 4th-order corrections enhance the two peaks.
Finally the ATA correction (the best result in the present work) reduces
the peaks and shifts them towards the Fermi level ($i.e.$, the low
energy side).

We have reported in our recent paper~\cite{kake10} that the 4th-order 
dynamical
corrections improve the magnetic properties of Fe and Ni.  We obtained
the Curie temperature $T_{\rm C}=2070$ K (1420 K) for Fe (Ni) in the
static approximation.  The second-order dynamical corrections 
lead to $T_{\rm C}= 2020$ K (1260 K) for Fe (Ni).  The 4th-order
dynamical corrections further reduce $T_{\rm C}$ to 1930 K for Fe 
and 620 K for Ni, respectively.  
The latter is in good agreement with the experimental 
value 630 K, while the former is still overestimated by a factor
of 1.8.  A large reduction of $T_{\rm C}$ in Ni due to the 4th-order 
dynamical corrections has been attributed to a
reduction of the DOS at the Fermi level.  
Calculated effective Bohr magneton numbers in the
2nd-order dynamical CPA are 3.0 $\mu_{\rm B}$ and 1.2 $\mu_{\rm B}$ for
Fe and Ni, respectively.  The 4th-order dynamical corrections yield 3.0 
$\mu_{\rm B}$ for Fe and 1.6 $\mu_{\rm B}$ for Ni, both of which are in
good agreement with the experimental values 3.2 $\mu_{\rm B}$ (Fe) and
1.6 $\mu_{\rm B}$ (Ni).
%
\begin{figure}
\includegraphics{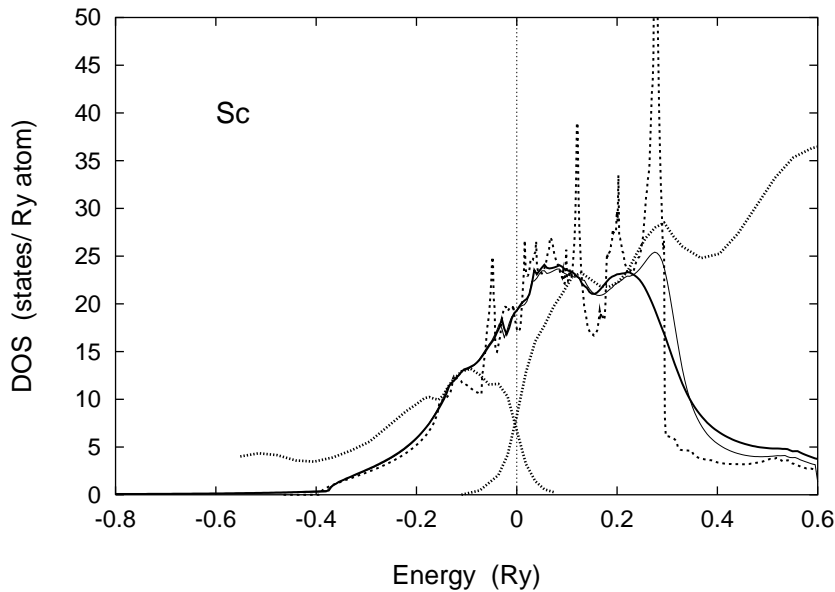}%
\caption{\label{figsc}
Densities of states (DOS) as the single-particle excitation spectra 
for fcc Sc.
DOS in the static approximation: thin solid curve, DOS with dynamical
correction: solid curve, DOS in the local density approximation (LDA):
dashed curve.
The X-ray photoemission spectra (XPS)~\cite{narm88} 
and inverse photoemission spectra (BIS)~\cite{spei84} 
for hcp Sc at room temperature are shown by dotted curves.
Note that these data are arbitrarily renormalized in order to fit the
calculated DOS.
}
\end{figure}
%

Among the 3$d$ transition metals, scandium has the weakest Coulomb
interaction as shown in Fig. 1.  Calculated DOS for fcc Sc are presented in
Fig. 3.  The DOS below the Fermi level is close to the LDA DOS except
some detailed structures.  This is due to a small 
number of $d$ electrons per orbital ($\sim 0.3$ per $d$
orbital) and rather weak Coulomb interactions.  
It should be noted that there is no correlation correction to the 
$sp$ bands in the present approximation (\ref{cohg2}), 
so that spiky structure of the $sp$ bands in the LDA calculations 
remains in the total DOS.   
The $d$ DOS are smoothed by the scattering corrections of the 
self-energy.  The corrections become larger near the top of 
$d$ bands, so that the peak of t${}_{\rm 2g}$ band at $\omega = 0.275$ Ry 
is much damped down, the $d$ band becomes narrower, and
the spectral weight shifts to the higher energy region.  The
difference between the dynamical and static DOS is rather small except 
high energy region ($\omega \gtrsim 0.2$ Ry).
We performed the numerical calculations with use of the screened
value $\overline{J}=0.029$ Ry, but we hardly found the change of DOS 
in shape.

%
\begin{figure}
\includegraphics{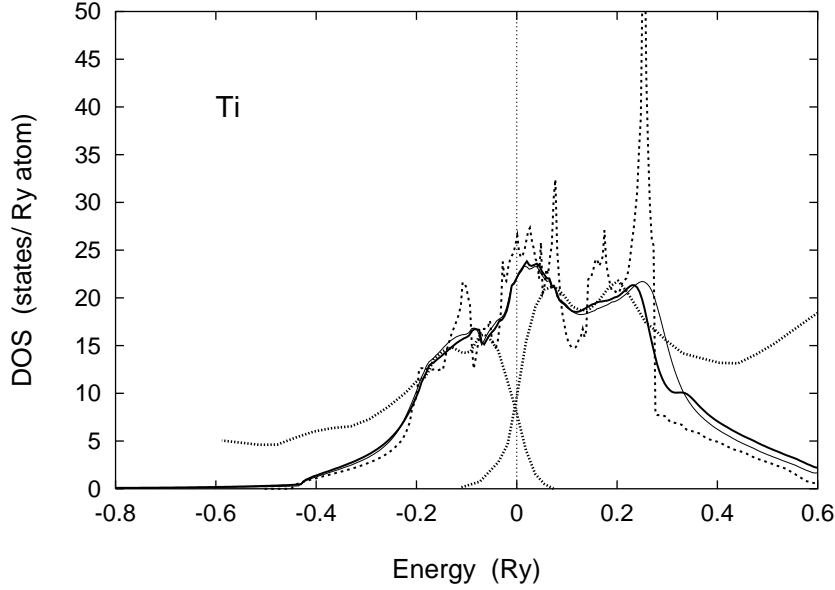}%
\caption{\label{figti}
Calculated DOS for fcc Ti.
The notations are the same as in Fig. 3.
The XPS~\cite{narm88} and BIS~\cite{spei84} data 
for the hcp Ti are obtained at room temperature.
}
\end{figure}
%
Calculated DOS qualitatively agree with the
XPS and BIS data for hcp Sc~\cite{narm88,spei84} as shown in Fig. 3. 
(Note that the crystal structure of Sc is
not the fcc but the hcp experimentally.)
Here and in the followings the intensities of the experimental data are
arbitrarily scaled to fit theoretical DOS.
Rapid decrease of the XPS and BIS data indicates the cut-off due to
Fermi distribution function.  Moreover the deviations from the DOS in high
energy region are due to secondary electrons, and outside the scope of
the present theory.
The high-energy peak in the calculated DOS around $\omega = 0.25$ Ry 
deviates from the BIS peak at $\omega = 0.30$ Ry.
This is partly explained by the difference in crystal structure.
In fact, the LDA calculations~\cite{blah88} 
indicate that the peak position in the hcp Sc is
higher than that of the fcc one by 0.025 Ry.
A small hump at about $\omega = -0.2$ Ry in the XPS data does not
appear in the present calculations.  The discrepancy is not due to the
hcp crystal structure since there is no corresponding peak in the 
LDA DOS for the hcp Sc~\cite{blah88}.

In the case of fcc Ti, we obtained the DOS being similar to the fcc Sc
as shown in Fig. 4.
Thermal excitations smooth the LDA DOS, damp the highest peak at
$\omega = 0.25$ Ry, and transfer the spectral weight to the higher
energy region ($\omega \gtrsim 0.30$ Ry).  The dynamical corrections are
not so important in the case of Ti as understood by comparing the DOS 
with that in the static approximation.  
We find rather good agreement of the DOS with both the XPS and BIS 
experimental data for hcp Ti.
%
\begin{figure}
\includegraphics{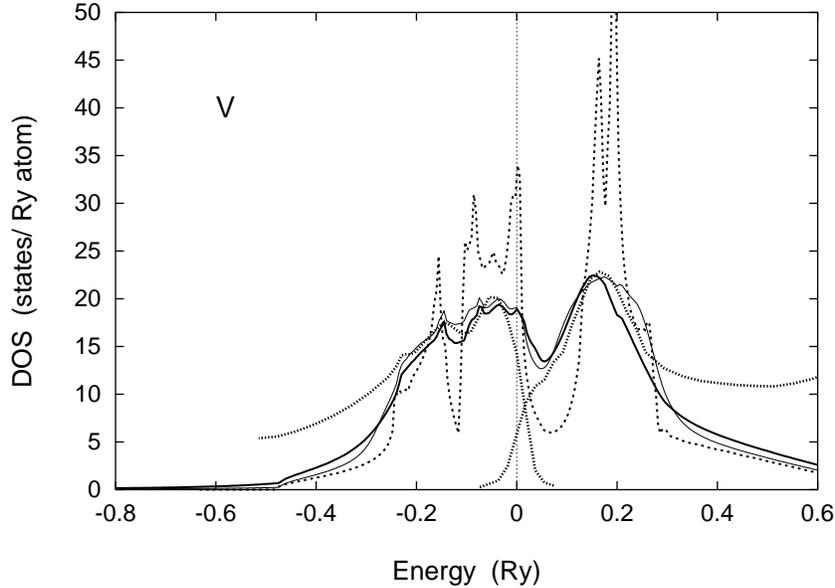}%
\caption{\label{figv}
Calculated DOS for bcc V.
The notations are the same as in Fig. 3.
The XPS~\cite{narm88} and BIS~\cite{spei84} data 
for the bcc V are obtained at room temperature.
}
\end{figure}
%

We have calculated the excitation spectra of vanadium for the bcc
structure as shown in Fig. 5.  
In this case the crystal structure is identical with 
the experimental one.  
The main peaks and valleys in the LDA DOS are much weakened by local 
electron correlations, and the spectral weights move to the higher 
energy region.  The $d$ bands in the quasiparticle energy region 
($|\omega| < 0.2$ Ry) shrink by about 10\% as compared with the LDA one.  
The calculated DOS shows a good agreement with the XPS and BIS 
data~\cite{narm88,spei84} in lineshape.  
Note that any artificial parameter 
is not introduced for comparison between the theory and 
experiment.  The DOS in the static approximation yields
an excess $d$ band broadening due to thermal spin fluctuations.  

Calculated excitation spectra of the bcc Cr is similar to that 
in the bcc V as shown in Fig. 6.  
Because the valence-electron number of Cr is larger
than that of V by one, the Fermi level shifts to the higher energy
region.  The t${}_{\rm 2g}$ peak around $\omega = -0.15$ Ry in the LDA DOS
is weakened, and shifts toward the Fermi level.
The position of the e${}_{\rm g}$ peak at $\omega = 0.1$ Ry is not
changed, but its weight is much decreased by electron correlations.  
The static approximation broadens the e${}_{\rm g}$ peaks excessively, 
thus does not explain the BIS data.
Calculated peak around $\omega = -0.1$ Ry seems to be too small as
compared with the XPS experiment.  There is a possibility that the
antiferromagnetic correlations enhance the peak in the present
calculations, because the experimental data are taken below the N\'{e}el
temperature. 
%
\begin{figure}
\includegraphics{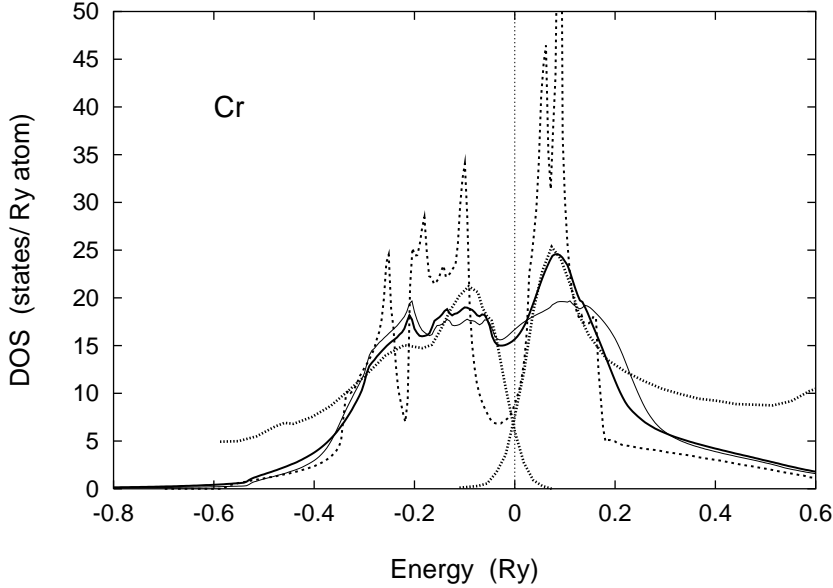}%
\caption{\label{figcr}
Calculated DOS for bcc Cr.
The notations are the same as in Fig. 3.
The XPS~\cite{narm88} and BIS~\cite{spei84} data 
for the bcc Cr are obtained at room temperature.
}
\end{figure}
%

The Coulomb interactions of Mn are roughly twice as large as those
in Sc, while their LDA DOS in the fcc structure 
are similar to each other.  In addition, electron 
number per atom increases from 3 to 7.  Thus one expects more
electron correlations in the case of Mn.  
Calculated DOS as well as experimental XPS-BIS
data~\cite{narm88, bier01} 
are shown in Fig. 7.  We find that the central peak consisting of the  
t${}_{\rm 2g}$ bands around $\omega = -0.1$ Ry in the LDA DOS changes to a
valley due to electron correlations, so that the DOS shows a two-peak
structure.  The result indicates a formation of the Mott-Hubbard type 
bands due to strong on-site correlations as we suggested in our recent 
paper~\cite{kake09}.  
The same result was obtained by Birmann {\it et. al.}~\cite{bier01} 
by using the Hamiltonian without transverse spin fluctuations. 
Static approximation overestimates the splitting of the Mott-Hubbard
bands.  The dynamical effects suppress such a band broadening due
to thermal spin fluctuations.
When we adopt the screened value $\overline{J}=0.043$ Ry, the DOS around
$\omega=-0.1$ Ry increases and the two-peak structure becomes less clear.
%
\begin{figure}
\includegraphics{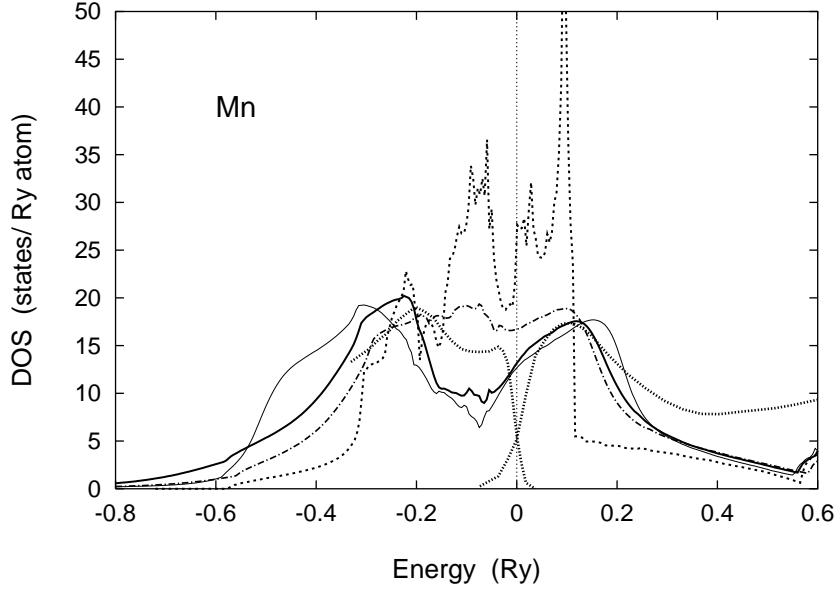}%
\caption{\label{figmn}
Calculated DOS for fcc Mn.
The notations are the same as in Fig. 3.
The dot-dashed curve is the dynamical results for $\overline{U}=0.192$
 Ry and the screened exchange energy parameter $\overline{J}=0.043$ Ry.
The XPS data~\cite{bier01} for the fcc thin-film Mn on 
Cu${}_{3}$Au(100) and the BIS data~\cite{spei84} 
for $\alpha$-Mn at room temperature are plotted for comparison with the
 theory.
}
\end{figure}
%

Experimentally, the bulk fcc Mn is realized only in the narrow temperature 
range between 1352 K and 1407 K at high temperatures.  
There is no photoemission experiment in this
temperature regime.  However the XPS data for the 20 monolayer fcc Mn on
Cu${}_{3}$Au(100)~\cite{bier01} are available at room temperature. 
The data seem to be explained by
the dynamical CPA with partially screened $\overline{J}$ between 0.043
and 0.061 Ry.  
Theoretical results agree with the BIS data~\cite{spei84} 
in peak position.

The line shape of the calculated DOS for bcc Fe is similar to the bcc Cr 
as shown in Fig. 8.  The main peak of the LDA DOS near the Fermi level 
and the peak of t${}_{\rm 2g}$
bands around $\omega = -0.15$ Ry are much weakened due to electron
correlations.  The spectral weight moves to higher energy region.  
The static approximation overestimates the band width by
20\%.  The main peak at $\omega = 0.04$ Ry 
above the Fermi level is consistent with
the BIS data~\cite{kirs84} 
at 0.86$T_{\rm C}$, $T_{\rm C}$ being the Curie
temperature.  (Note that a hump at $\omega = 0.1$ Ry in the BIS data is
the remnant of the e${}_{\rm 2g}$ peak for the minority band and should
disappear above $T_{\rm C}$.)
On the other hand, the DOS below the Fermi level does not well 
correspond to the XPS data~\cite{kirb85} 
at 1.03$T_{\rm C}$.  One needs more weight around $\omega =
-0.1$ Ry in order to explain the experimental data.  
%
\begin{figure}
\includegraphics{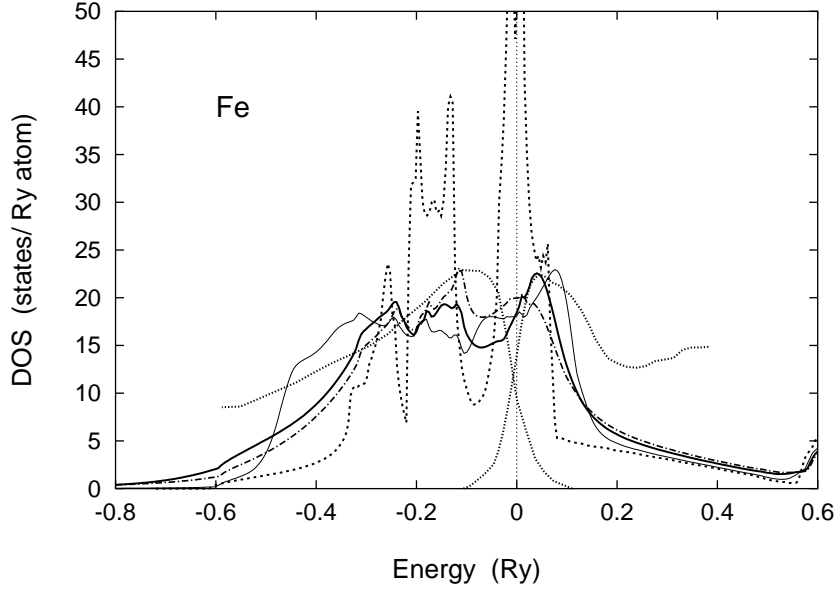}%
\caption{\label{figfe}
Calculated DOS for bcc Fe.
The notations are the same as in Fig. 3.
The dot-dashed curve is the dynamical results for $\overline{U}=0.169$
 Ry and the screened exchange energy parameter $\overline{J}=0.046$ Ry.
The XPS data~\cite{kirb85} are measured at 1.03$T_{\rm C}$, 
$T_{\rm C}$ being the Curie temperature.  The BIS 
data~\cite{kirs84} are measured at 0.86$T_{\rm C}$.
}
\end{figure}
%

The DOS calculated with use of the screened value $\overline{J}=0.046$
Ry considerably enhance the weight around $\omega=-0.1$ Ry, thus the
screening on $\overline{J}$ partly explains the broad peak at
$\omega=-0.1$ Ry in the XPS data.  This feature does not change even if
we adopt $\overline{U}$ value obtained by Bandyopadhyav {\it et. al.}
The magnetic short-range order~\cite{kisk87} may also 
explain the discrepancy between the XPS data and the present result
based on the single-site approximation because the experimental data of
the XPS are taken near $T_{\rm C}$. 
%
\begin{figure}
\includegraphics{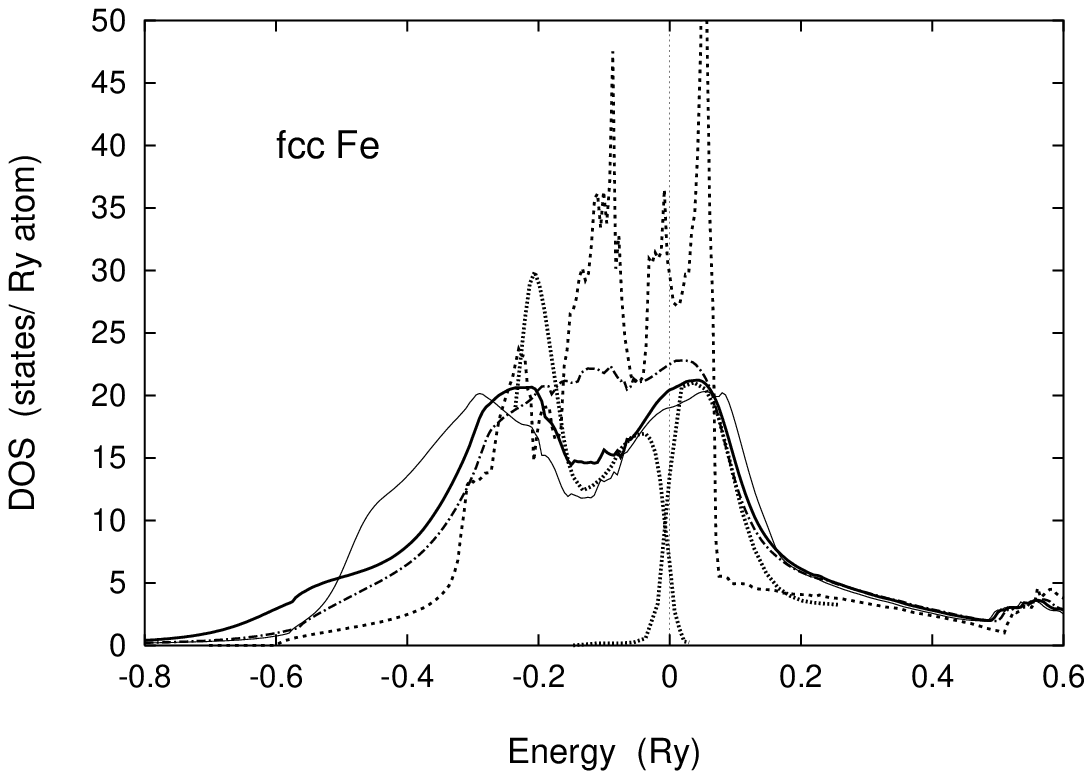}%
\caption{\label{figfccfe}
Calculated DOS for fcc Fe.
The notations are the same as in Fig. 3.
The dot-dashed curve is the dynamical results for $\overline{U}=0.169$
 Ry and the screened exchange energy parameter $\overline{J}=0.046$ Ry.
The XPS~\cite{zhar97} and BIS~\cite{himp91} data are obtained for 
the fcc Fe on Cu(100) at room temperature.
}
\end{figure}
%

We have also investigated the DOS for the fcc Fe.
The fcc Fe is well-known to be a typical itinerant magnet showing the spin
density waves with magnetic moment of about 1 $\mu_{\rm B}$ per atom at
low temperatures~\cite{tsuno89,kake02-3,uchi06}.  
But the fcc Fe shows anomalous thermal
expansion~\cite{chev28}, 
so that the $d$ band width at high temperatures is expected to become
narrower than those at low temperatures by several percent.  
We may then expect stronger correlation effects.  
As shown in Fig. 9, we find that 
the DOS calculated with use of the unscreened value $\overline{J}=0.066$
Ry is similar to that of the fcc Mn.  The t${}_{\rm 2g}$ central
peak in the LDA DOS splits into the lower and upper Mott-Hubbard
bands due to on-site correlations.  
The band splittings are smaller than those in the fcc Mn, 
and the dip at $\omega = -0.15$ Ry is weakened (see Fig. 7).
The static approximation overestimates the band width by 20 \%.

We present in Fig. 10 the $d$ partial DOS for e${}_{\rm g}$ and t${}_{\rm 2g}$
bands in order to examine the details of the band splitting.
As seen from the figure, both the e${}_{\rm g}$ and the t${}_{\rm 2g}$ DOS
show the two-peak structure.  The energy difference between the upper
and lower peaks is about 0.25 Ry, which is approximately equal to
the intra-orbital Coulomb interaction for fcc Fe, $U_{0} = 0.27$ Ry.  
Moreover, we find a large scattering peak of 
$- {\rm Im} \Sigma_{L\sigma}(\omega+i\delta)$ at $\omega=-0.12$ Ry. 
These behaviors verify that the
two-peak DOS forms the Mott-Hubbard bands due to electron correlations.
%
%
\begin{figure}
\includegraphics{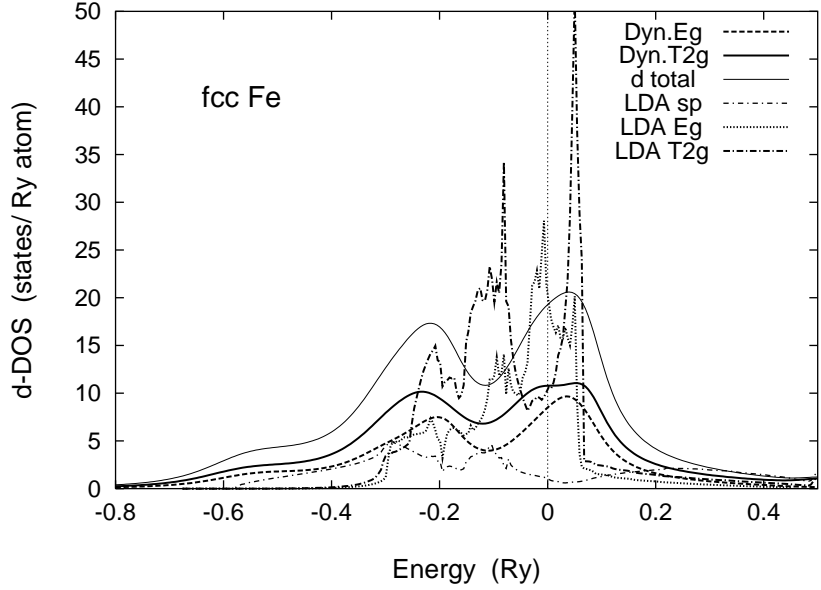}%
\caption{\label{figfccfed}
Partial $d$ DOS of fcc Fe for e${}_{\rm g}$ (dashed curve) 
and t${}_{\rm 2g}$ (solid curve) electrons.  The total $d$ DOS 
is shown by thin solid curve. The partial DOS in the LDA are also shown 
by dotted curve (e${}_{\rm g}$), dot-dashed curve (t${}_{\rm 2g}$),
and thin dot-dashed curve (sp electrons), respectively. 
}
\end{figure}
%
%

It is remarkable that the dynamical results are sensitive to the 
choice of $\overline{U}$ and $\overline{J}$ in the case of fcc Fe.  
The two-peak structure almost disappears when we adopt 
$\overline{U}=0.169$ Ry and the screened value $\overline{J}=0.046$ Ry, 
as shown in Fig. 9, while it again appears when we adopt 
$\overline{U}=0.219$ Ry obtained by Bandhyopadhyav {it et. al.} 
and the screened value $\overline{J}=0.046$ Ry.
The behaviors are understood from the following arguments.
The Coulomb interactions suppress the doubly occupied states of
electrons on an orbital in general. 
The Hund-rule coupling $\overline{J}$ also suppresses the
doubly occupied states to reduce the energy; 
$\overline{J}$ tends to enhance $\overline{U}$
effectively.  Thus the increment of $\overline{U}$ or $\overline{J}$ is
favorable for the formation of the Mott-Hubbard bands. 

The peak near the Fermi level explains well the BIS data~\cite{himp91} 
for the fcc Fe
on Cu(100).  The XPS data~\cite{zhar97} 
for the fcc Fe on Cu(100) are somewhat
controversial.  The peak around $\omega = -0.2$ Ry is usually
interpreted as a peak due to Cu substrate.  If this peak originates in
the Cu substrate by 100\%, the data support the result for
$(\overline{U}, \overline{J})=(0.169, 0.046)$ Ry.
But if the peak is interpreted as a superposition of both the
fcc Fe and the Cu substrate spectra, 
we expect a two-peak structure of fcc Fe,
and the XPS data are consistent with the results for 
$(\overline{U}, \overline{J})=(0.169, 0.066)$ and 
$(\overline{U}, \overline{J})=(0.219, 0.046)$ Ry.
Resolving the problem is left for future investigations.
%
%
\begin{figure}
\includegraphics{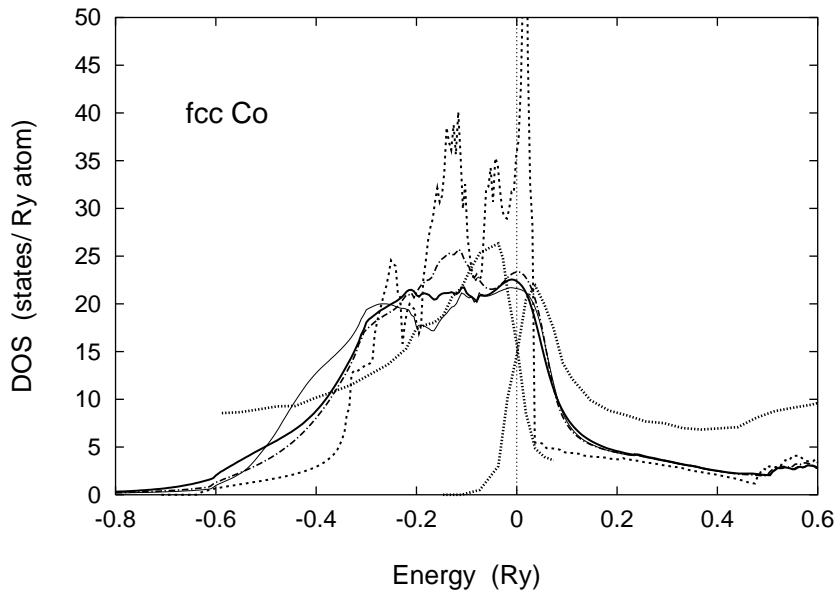}%
\caption{\label{figco}
Calculated DOS for fcc Co.
The notations are the same as in Fig. 3.
The dot-dashed curve is the dynamical results for $\overline{U}=0.245$
 Ry and the screened exchange energy parameter $\overline{J}=0.048$ Ry.
The XPS~\cite{narm88} and BIS~\cite{spei84} data for the hcp Co 
at room temperature are drawn by the dotted curves.
}
\end{figure}
%
%

The DOS of fcc Co in the paramagnetic state does not show the
Mott-Hubbard type structure any more as shown in Fig. 11.  
The peak of the t${}_{\rm 2g}$ bands at $\omega = -0.15$ Ry becomes almost
flat, and the peak of the t${}_{\rm 2g}$ bands on the Fermi 
level is much weakened. 
We find a weak hump around $\omega = -0.5$ Ry suggesting the `6 eV'
satellite. 
When we adopt the screened value $\overline{J}=0.048$ Ry, the peak 
at $\omega=-0.15$ Ry partially remains as found in Fig. 11, 
and the hump around $\omega=-0.5$ Ry almost disappear.

We present in Fig. 12 the e${}_{\rm g}$ and t${}_{\rm 2g}$ partial DOS 
to clarify the disappearance of the two-peak
structure in the total DOS.  Calculated t${}_{\rm 2g}$ partial DOS does not
show the band splitting any more.  The e${}_{\rm g}$ DOS still show the
two-peak structure.  But the dip between the peaks is shallower 
than that of the fcc Fe (see Fig. 10) and
the energy difference between the peaks (0.18 Ry) is smaller than the
intra-orbital Coulomb interaction energy $U_{0} = 0.36$ Ry.
In the case of Co, the number of hole states is reduced as compared with
that of fcc Fe.  This reduces the charge fluctuations and thus the
magnitude of the self-energy by a factor of two.  The reduction yields the
disappearance of the two-peak structure in the total DOS.
%
%
\begin{figure}
\includegraphics{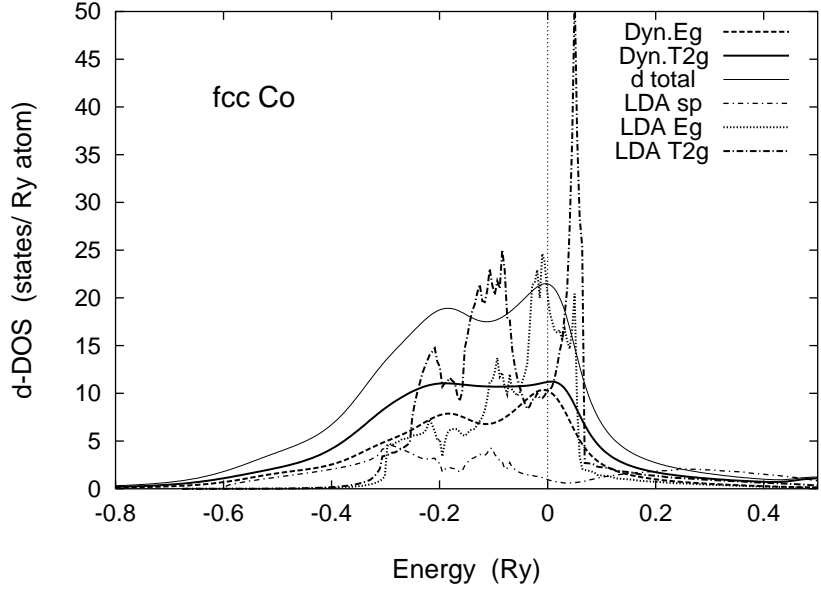}%
\caption{\label{figfccfed}
Partial $d$ DOS of fcc Co for e${}_{\rm g}$ (dashed curve) 
and t${}_{\rm 2g}$ (solid curve) electrons.  The total $d$ DOS 
is shown by thin solid curves. The partial DOS in the LDA are also shown 
by dotted curve (e${}_{\rm g}$), dot-dashed curve (t${}_{\rm 2g}$),
and thin dot-dashed curve (sp electrons). 
}
\end{figure}
%
%

There is no experimental data on the fcc Co above the Curie temperature.
Calculated DOS does not show a good agreement with the XPS~\cite{narm88}
and BIS data~\cite{spei84} of hcp Co at room temperature.  
The UPS (Ultraviolet Photoemission Spectroscopy) data for hcp Co by 
Heimann {\it et. al.}~\cite{heim77} 
show the existence of the 6 eV satellite in agreement with the
present result for the unscreened $\overline{J}$, while the other 
data~\cite{narm88, huf74} do not.

Single-particle excitations of Ni have been investigated extensively 
in both theory~\cite{penn79,lieb79,vict85,unger94,lich01,kake08} 
and experiment~\cite{east78,himp79,east80,eber80,mar84}.  
Present result of fcc Ni shows a single-peak structure as shown in Fig. 13.  
Moreover the correlations increase the spectral weight
around $\omega = -0.45$ Ry, and creates a small hump
corresponding to the 6 eV satellite due to two-hole excitations.  
The dynamical effects suppress the band broadening of
the static approximation by 20 \%.  
These behaviors do not change even if we adopt the screened value
$\overline{J}=0.046$ Ry.  The screened $\overline{J}$ enhances the main
peak around $\omega=-0.05$ Ry and creates a hump at $\omega=-0.15$ Ry
in the DOS.
Though the calculated DOS is consistent with
the XPS~\cite{narm88} and BIS data~\cite{spei84}, 
the band width seems to be somewhat larger than that of the XPS data.

We present finally in Fig. 14 the DOS for Cu.  Electron correlations via
a small number of $d$ holes ($\approx 0.36$ per atom) move
the spectral weight of the LDA DOS to the lower energy region.  
The peak of the $d$ bands
shifts toward lower energy by 0.05 Ry.  It is also remarkable that
a broad hump appears at $\omega = -0.7$ Ry due to
two-hole excitations.  We find a good agreement between the dynamical
CPA theory and the experiment~\cite{narm88,spei84} 
for this system.
%
\begin{figure}
\includegraphics{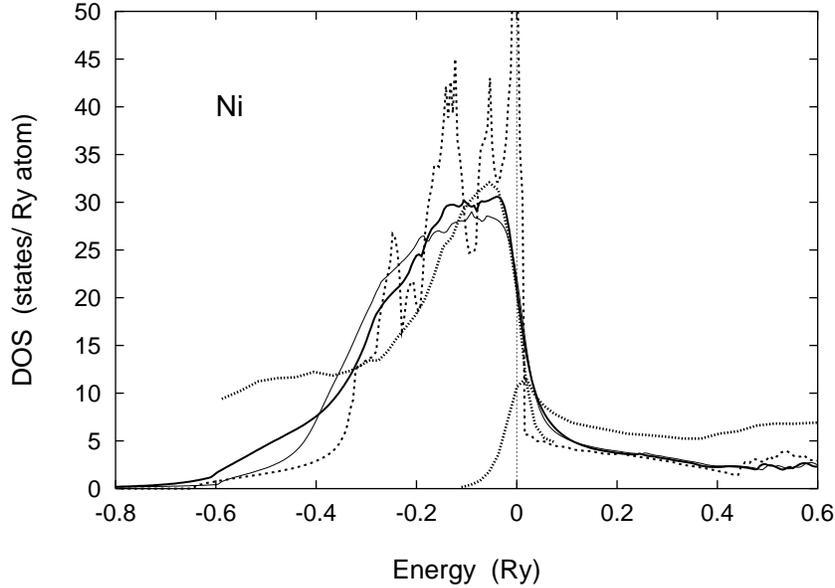}%
\caption{\label{figni}
Calculated DOS for fcc Ni.
The notations are the same as in Fig. 3.
The XPS~\cite{narm88} and BIS~\cite{spei84} data for the fcc Ni 
are obtained at room temperature.
}
\end{figure}
%

\section{Summary}

In the present paper, we have obtained approximate expression of the 
higher-order dynamical
corrections to the effective potential 
in the first principles dynamical CPA, 
making use of an asymptotic approximation.  
The approximation becomes exact in the high frequency limit, and much
reduces the multiple summations with respect to the Matsubara frequency
at each order of expansion in the dynamical corrections.
%
\begin{figure}
\includegraphics{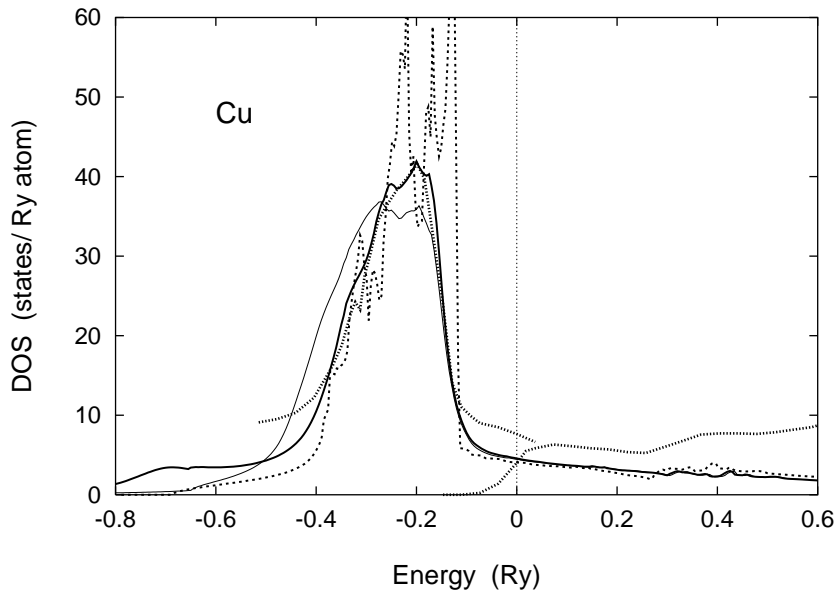}%
\caption{\label{figcu}
Calculated DOS for fcc Cu.
The notations are the same as in Fig. 3.
The XPS~\cite{narm88} and BIS~\cite{spei84} data for the fcc Cu 
are obtained at room temperature.
}
\end{figure}
%

Within the 4-th order dynamical corrections, we have investigated
systematic change of the DOS at high temperatures in 3$d$ series 
from Sc to Cu.   
Thermal spin fluctuations in the static approximation 
smooth the LDA DOS at high temperatures, 
especially reduce their main peaks and broaden the $d$ band width.  
Dynamical effects reduce the band broadening, 
and move the spectral weight to higher energy region. 
These effects explain in many cases the lineshape of the XPS and BIS 
experimental data from 
Sc to Cu quantitatively or semiquantitatively.
We found the formation of the Mott-Hubbard type bands due to electron 
correlations in the case of fcc Mn and fcc Fe,
and also found that the dynamical effects can create a small hump 
corresponding to `6 eV' satellite in Co, Ni, and Cu.

We investigated the effects of the screened exchange energy
parameters using the reduced values of $\overline{J}$ by 30 \%.  
The screening of $\overline{J}$ is significant for the DOS in
Mn, Fe, and Co.  The reduction of $\overline{J}$ tends to weaken the 
Mott-Hubbard type bands in fcc Mn, and even could destroy the two-peak
structure in the case of fcc Fe.  It also develops the central peak
around $\omega=-0.1$ Ry in bcc Fe and fcc Co.  Some of these results
explain better the XPS data.  But we have to calculate the other physical
quantities with use of the same scheme, and have to examine in more details
the consistency among them in order to conclude the validity of the
screened values of $\overline{J}$.  The magnetic short-range 
order should also be important for understanding the experimental 
data of magnetic transition metals for more detailed 
discussions.

Present theory explains a systematic change of the spectra in 3$d$
series at high temperatures.  At lower temperatures, the higher-order 
dynamical corrections should be more important.  Improvements of 
the theory in the low temperature region are left for future 
investigations.

\begin{acknowledgments}

Numerical calculations have been partly carried out with use of the
Hitachi SR11000 in the Supercomputer Center, Institute of Solid State
Physics, University of Tokyo.

\end{acknowledgments}


\end{document}